\documentclass[10pt]{article}
\usepackage[T1]{fontenc}
\usepackage[utf8]{inputenc}
\usepackage{textcomp}
\usepackage{amsmath}
\usepackage{amsfonts}
\usepackage{graphicx}
\usepackage{color}
\usepackage{amsfonts}
\usepackage{amssymb}
\usepackage{mathrsfs}
\usepackage[table,xcdraw]{xcolor}
\usepackage{orcidlink}
\usepackage{hyperref}
\usepackage{etoolbox}     
\usepackage{orcidlink}
\hypersetup{colorlinks=true, linkcolor=blue, citecolor=blue, urlcolor=blue}
\usepackage{url}
\usepackage{float}
\usepackage{multirow}
\usepackage{array}
\usepackage{longtable}
\usepackage{booktabs}
\usepackage{colortbl}
\usepackage{subcaption}
\usepackage{geometry}
\usepackage[numbers,sort&compress]{natbib}
\begin{document}
\baselineskip=20pt
\begin{center}
\LARGE{Thermal and Optical Signatures of Einstein-Dyonic ModMax Black Holes with GUP and Plasma Modifications}
\end{center}
\vspace{0.3cm}
\begin{center}
{\bf Erdem Sucu \orcidlink{0009-0000-3619-1492}}\footnote{\bf erdemsc07@gmail.com}\\
{\it Eastern Mediterranean University, Physics Department, Famagusta, 99628 North Cyprus, via Mersin 10, T\"urkiye}\\
{\bf Suat Dengiz \orcidlink{0000-0002-7099-4608}}\footnote{\bf suat.dengiz@atilim.edu.tr}\\
{\it Department of Electrical and Electronics Engineering, Faculty of Engineering, At{\i}l{\i}m University, 06836 Ankara, T\"urkiye}\\
{\bf \.{I}zzet Sakall{\i} \orcidlink{0000-0001-7827-9476}}\footnote{\bf izzet.sakalli@emu.edu.tr (Corresp. author)}\\
{\it Eastern Mediterranean University, Physics Department, Famagusta, 99628 North Cyprus, via Mersin 10, T\"urkiye}
\end{center}
\vspace{0.3cm}
\abstract{We explore the thermodynamic and optical properties of Einstein-Dyonic-ModMax (EDM) black holes (BHs) incorporating quantum gravity corrections and plasma effects. The ModMax theory promotes the classical Maxwell theory to a non-linear electrodynamics with a larger symmetry structure (electromagnetic duality plus conformal invariance), and provides dyonic BH solutions characterized by both electric and magnetic charges modulated by the nonlinearity parameter $\gamma$. Using the Hamilton-Jacobi tunneling formalism, we derive the Hawking radiation spectrum and demonstrate how the Generalized Uncertainty Principle (GUP) modifies the thermal emission, potentially leading to stable remnants. Our analysis of gravitational lensing employs the Gauss-Bonnet theorem to compute light deflection angles in both vacuum and plasma environments, revealing strong dependencies on the ModMax parameter and plasma density. We extend this to axion-plasmon environments, uncovering frequency-dependent modifications that could serve as dark matter signatures. The photon motion analysis in plasma media shows how the exponential damping term $e^{-\gamma}$ affects electromagnetic backreaction on spacetime geometry. We compute quantum-corrected thermodynamic quantities, including internal energy, Helmholtz free energy, pressure, and heat capacity, using exponentially modified entropy models. The heat capacity exhibits second-order phase transitions with critical points shifting as functions of $\gamma$, indicating rich thermodynamic phase structures. The energy condition analysis shows that classical ModMax electrodynamics satisfies the null and weak energy conditions, while the observed near-horizon violations arise only after incorporating quantum-corrected entropy effects.}

\vspace{0.3cm}

\section{Introduction} \label{iz1}

BHs stand as profound arenas where the geometrical elegance of general relativity intertwines with the complexities of quantum physics and high-energy field theories \cite{giddings2019black,goldberger2020effective}. In classical settings, the Einstein-Maxwell framework provides well-known charged BH solutions, such as the Reissner-Nordstr\"{o}m metric. However, this linear theory fails to capture nonlinear electromagnetic interactions that may become significant in extreme gravitational or strong-field regimes, such as near compact objects (including accretion disks surrounding BHs) or early-universe conditions, or they might even further supply viable seeds for BH singularities. In this sense, as is well known, the Born-Infeld \cite{Born:1934gh} and the Euler-Heisenberg \cite{Heisenberg:1936nmg} nonlinear models of electrodynamics in the literature come forward as pioneering frameworks for understanding electromagnetic phenomena beyond the linear regime. (Also, see, e.g., \cite{Dengiz:2011ig} for an interesting alternative modification of Born-Infeld theory in Weyl's geometry.)

Recently, the nonlinear extensions of electrodynamics have gained renewed interest, notably the Modified Maxwell (ModMax) theory \cite{Bandos:2020jsw, Kosyakov:2020wxv,baptista2025scattering}, which generalizes Maxwell electrodynamics while preserving duality and conformal invariance in vacuum and thus has a larger symmetry structure. Recent progress has significantly expanded the landscape of ModMax BH solutions in various gravitational frameworks. For instance, exact analytical BH solutions have been obtained in F(R)-ModMax gravity, revealing nontrivial thermodynamic structures and stability conditions~\cite{EslamPanah:2024tex}. Moreover, extensions to massive gravity and modified theories have led to novel ModMax-dRGT-like BH configurations with rich phase structures~\cite{EslamPanah:2025bfh}. Very recent developments have also explored dyonic ModMax BHs in alternative gravity sectors such as Kalb--Ramond theory, where geodesics, shadows, and Hawking radiation were analyzed~\cite{Ahmed:2026vbh,Ali:2025iyl}. These results demonstrate the rapidly growing relevance of ModMax electrodynamics in modern BH physics. The minimal coupling of ModMax theory to Einstein gravity leads to the so-called EDM BHs, offering rich structure due to the interplay between nonlinear electromagnetic (NLE) fields and curved spacetime \cite{Bokulic:2025usc,sahan2025quantum,kruglov2022magnetic,Zubair:2023cor,Mazharimousavi2022}. Recent investigations have extended these studies to Kalb-Ramond ModMax BHs, where the inclusion of antisymmetric tensor fields and Lorentz symmetry breaking parameters creates additional modifications to particle dynamics and thermal properties \cite{Ahmed:2025bpg}. This theoretical framework represents a significant advancement in our understanding of how fundamental electromagnetic nonlinearities manifest in gravitational contexts, providing a bridge between quantum field theory predictions and classical general relativistic descriptions.

A particularly interesting subclass of these solutions emerges when both electric and magnetic charges are present, forming dyonic BHs \cite{Chow:2013gba,Kruglov:2020aqm}. These objects are characterized not only by their mass and horizon structure but also by a duality-rotated electromagnetic field, wherein the ModMax parameter $\gamma$ governs the strength of the nonlinearity \cite{pantig2022shadow,Bokulic:2025usc}. The inclusion of this parameter significantly modifies the causal structure, the effective potential experienced by test particles and photons, and the thermodynamic quantities defined on the horizon. Importantly, the ModMax-induced exponential damping of the electromagnetic terms enables a smooth interpolation between nonlinear electrodynamics and the standard Einstein-Maxwell theory, providing a powerful model to probe beyond-linear interactions in curved backgrounds \cite{Bandos:2021rqy,Ahmed:2025rby}. In the present work, our primary objective is to investigate how the nonlinear ModMax parameter and quantum gravity corrections influence both the thermal and optical observables of Einstein--Dyonic ModMax BHs. In particular, we analyze the Hawking radiation spectrum via the Hamilton--Jacobi tunneling method, incorporate Generalized Uncertainty Principle (GUP) corrections \cite{pedram2011effects,ali2025first,sucu2026multi,Sucu:2025pce}, and study photon trajectories and gravitational lensing in vacuum and plasma environments. It is worth emphasizing that the GUP has been extensively developed within a broad range of quantum gravity approaches, including string theory, loop quantum gravity, and doubly special relativity frameworks. Foundational formulations of the GUP were introduced in the seminal works of Maggiore~\cite{maggiore1993generalized}, Scardigli~\cite{Scardigli:1999jh}, and Adler \textit{et al.}~\cite{adler2001generalized}, while later refinements incorporating linear and higher-order momentum corrections were proposed by Ali \textit{et al.}~\cite{Ali:2009zq}. More recent developments have explored the phenomenological implications of GUP in BH physics, including modifications to Hawking radiation, BH remnants, and observational signatures, as discussed in~\cite{Nozari:2005ex,ong2018generalized}. These different analyses are not independent components; rather, they provide complementary perspectives on how the ModMax nonlinearity parameter $\gamma$ affects observable properties of the spacetime \cite{Gibbons:1995cv}. The tunneling analysis determines the quantum-corrected temperature, the optical analysis probes photon propagation and lensing signatures, and the thermodynamic study clarifies how these effects are reflected in horizon thermodynamics.

The parameter $\gamma$ acts as a tuning knob that controls the departure from Maxwell theory, allowing investigations of how nonlinear electromagnetic effects influence BH properties across different energy scales \cite{EslamPanah:2024tex,Shahzad:2024ljt}. Unlike purely electric or magnetic solutions, dyonic configurations exhibit electromagnetic duality symmetries that are preserved under the ModMax transformations, leading to novel theoretical predictions and potential observational signatures \cite{ayon2024nonlinearly,barrientos2025electromagnetized}.

From a semi-classical perspective, BHs radiate thermally through the mechanism of Hawking radiation. In this study, the tunneling formalism \cite{gecim2013tunnelling}, specifically the Hamilton-Jacobi approach, is employed to derive the thermal spectrum associated with EDM BHs. Furthermore, quantum gravity corrections are incorporated through the GUP \cite{ramezani2024linear,rizwan2025gup,bhandari2025generalized,gecim2020quantum,Sakalli:2025els,Ali2009GUP}, which encodes the existence of a minimal length scale expected from various quantum gravity candidates. These corrections not only alter the BH's evaporation profile but also suggest the emergence of remnants, as the corrected Hawking temperature ceases to diverge. Such remnants, if stable, could serve as potential dark matter candidates or markers of Planck-scale physics in BH environments \cite{khodadi2018planck,NooriGashti:2025xfg,Nozari:2005ex,ong2018generalized}.

In addition to their thermal properties, BHs act as natural gravitational lenses. The propagation of photons in EDM backgrounds is strongly influenced by both the nonlinear electromagnetic charges and the medium through which they travel. By embedding these spacetimes in dispersive media such as plasma and extending further to axion-plasmon environments, this study investigates the deflection of light using the Gauss-Bonnet theorem (GBT) in its optical geometry formulation \cite{sucu2025charged,ovgun2019deflection,Sucu_2026}. The analysis reveals a sensitive dependence of the deflection angle on the ModMax parameter $\gamma$, plasma frequency, and impact parameter \cite{Guzman-Herrera:2023zsv}.

The gravitational lensing analysis extends beyond traditional vacuum studies by incorporating realistic astrophysical environments where BHs are embedded in complex plasma structures. Such environments are ubiquitous in nature, from accretion disks around stellar-mass BHs to the magnetized plasma surrounding supermassive BHs in galactic centers \cite{bisnovatyi2017gravitational,ibrokhimov2025testing,kumar2024shadow,Perlick2022,PerlickTsupko2015}.

The thermodynamic structure of EDM BHs is revisited under quantum-corrected entropy models that incorporate exponential modifications beyond the Bekenstein-Hawking area law \cite{chatterjee2020exponential,gursel2025thermodynamics,sucu2025quantum1,soroushfar2023non}. The corrected forms of internal energy, Helmholtz free energy, pressure, and heat capacity are computed, demonstrating phase transition behavior that is strongly regulated by the nonlinearity parameter $\gamma$. Furthermore, this investigation addresses fundamental questions about energy conditions and the physical viability of EDM spacetimes \cite{visser1995lorentzian,campos2022quantum,sekhmani2025thermodynamics}.

Among the parameters entering the model, the ModMax nonlinearity parameter $\gamma$ plays a central role in shaping both the optical and thermodynamic signatures of the BH. In particular, the deflection angle and photon sphere radius exhibit a pronounced sensitivity to $\gamma$, suggesting that future horizon-scale observations, such as those provided by the Event Horizon Telescope, could potentially place observational bounds on this parameter in strongly magnetized environments. It is worth emphasizing that current horizon-scale observations probe photon trajectories near the photon sphere and the associated shadow size. In realistic astrophysical situations the deviations induced by nonlinear electrodynamics are expected to remain small unless the parameter $\gamma$ attains sufficiently large values. Therefore, the present analysis should be interpreted primarily as a theoretical investigation; nevertheless, future high-precision measurements of photon rings and shadow structures may provide indirect constraints on nonlinear electrodynamics parameters such as $\gamma$ \cite{EHT2019L1,EHT2022L12,Ayzenberg:2023hfw}.

The paper is organized as follows: Section~\ref{iz2} establishes the fundamental framework of EDM theory and examines the resulting BH solutions. Section~\ref{iz3} analyzes Hawking radiation via quantum tunneling methods. Section~\ref{iz4} incorporates GUP corrections to scalar field tunneling. Sections~\ref{iz5} and~\ref{iz6} investigate light deflection in plasma environments and vacuum using the GBT framework. Section~\ref{iz7} explores deflection in axion-plasmon environments. Section~\ref{iz8} examines photon motion in ModMax BHs. Section~\ref{iz9} analyzes quantum effects on BH thermodynamics. Section~\ref{iz10} investigates energy conditions and stress-energy tensor properties. Finally, Section~\ref{iz11} presents our conclusions and discusses future research directions.

\section{Fundamental Features of BHs in the Einstein-Dyonic-ModMax Framework} \label{iz2}

In this section, we establish the fundamental framework of the EDM theory and examine the resulting BH solutions. The most general Einstein-NLE action, which possesses an enriched symmetry structure including $SO(2)$ and conformal symmetries alongside ordinary diffeomorphisms, takes the form \cite{Bandos:2020jsw, Kosyakov:2020wxv}\footnote{Note that one can also assume a cosmological constant term and proceed accordingly.}:
\begin{equation}
    I=\frac{1}{16\pi} \int d^4 x \sqrt{-g} \, \Big [R+4  \mathcal{L}(\mathscr{F}, \mathscr{G}) \Big].
    \label{actionmainn}
\end{equation}
The NLE Lagrangian $\mathcal{L}(\mathscr{F}, \mathscr{G})$ is constructed from Lorentz and parity-invariant scalars built from the abelian field strength tensor $F_{\mu\nu}=\partial_\mu A_\nu-\partial_\nu A_\mu$ and its Hodge dual $\star F_{\mu\nu}$:
\begin{equation}
     \mathscr{F}= F_{\mu\nu}F^{\mu\nu} \quad \text{and} \quad  \mathscr{G}= F_{\mu\nu}\star F^{\mu\nu},
\end{equation}
where the Hodge star operation is defined as:
\begin{equation}
    \star F_{\mu\nu}=\frac{1}{2} \epsilon_{\mu\nu}{^{\alpha\beta}} F_{\alpha\beta}.
\end{equation}
Variation of Eq.~(\ref{actionmainn}) with respect to $g^{\mu\nu}$ and $A^\mu$ yields the metric and gauge field equations as follows:
\begin{equation}
    G_{\mu\nu}=8\pi \mathcal{T}_{\mu\nu}, \quad  d \mathbf{F}=0, \quad d \star \mathbf{Z}=0,
    \label{fieldequations}
\end{equation}
where $\mathbf{Z}=-4 (\mathcal{L}_{\mathscr{F}} \mathbf{F}+\mathcal{L}_{\mathscr{G}} \star \mathbf{F})$ with the abbreviations of $\mathcal{L}_{\mathscr{F}}=\partial_\mathscr{F} \mathcal{L}$ and $\mathcal{L}_{\mathscr{G}}=\partial_{\mathscr{G}} \mathcal{L}$. Also, $G_{\mu\nu}$ is the bare Einstein tensor and ${\cal T}_{\mu\nu}$ is the emerging stress-energy tensor. Moreover, the conserved electric and magnetic charges corresponding to the NLE sector can be evaluated through the prominent Komar integrals as:
\begin{equation}
   Q = \frac{1}{4\pi}\oint_{\hat{\Sigma}} \star \mathbf{Z} \quad \text{and} \quad  P = \frac{1}{4\pi}\oint_{\hat{\Sigma}} \mathbf{F},
   \label{komarint}
\end{equation}
where $\hat{\Sigma}$ represents a 2-dimensional closed surface. 

As to the particular NLE model of ModMax electrodynamics, the Lagrangian exclusively takes the ensuing form \cite{Bokulic:2025usc}:
\begin{equation}
    \mathcal{L}^{MM}=\frac{1}{4} \Big(-\mathscr{F} \cosh \gamma +\sqrt{ \mathscr{F}^2+ \mathscr{G}^2} \sinh\gamma \Big).
    \label{modmaxlag}
\end{equation}
Here $\gamma$ is a positive dimensionless parameter controlling the nonlinearity strength. Physically, the parameter $\gamma$ controls the deviation from the standard Maxwell electrodynamics. When $\gamma \rightarrow 0$, the theory smoothly reduces to the linear Einstein--Maxwell case, while nonzero values introduce nonlinear electromagnetic interactions that modify the effective stress--energy tensor sourcing the spacetime geometry. Consequently, $\gamma$ influences several observable properties of the BH, including the horizon structure, the effective photon potential, and the thermodynamic quantities associated with the event horizon. In the following sections we therefore treat $\gamma$ as the key parameter governing the deviations from standard charged BH behavior. This Lagrangian is self-dual and reduces to ordinary Maxwell theory when $\gamma \rightarrow 0$ \cite{Bandos:2020jsw, Kosyakov:2020wxv}. (See also \cite{Bokulic:2025usc, Barrientos:2022bzm, Bokulic:2021dtz, Flores-Alfonso:2020euz, Sorokin:2021tge, Bokulic:2022cyk, Nastase:2021ffq, Yasir:2025xgh, Ortaggio:2022ixh, Babaei-Aghbolagh:2022uij, Escobar:2021mpx, Shahzad:2024ljt} for a thorough understanding of Einstein-ModMax NLE.)

The ModMax theory enables the construction of dyonic BHs through electromagnetic duality rotations. More precisely, under the transformations:
\begin{equation}
\begin{aligned}
 \tilde{\mathbf{F}}&=\sin\alpha \, e^{\gamma} \star \mathbf{F}+\cos\alpha \, \mathbf{F},\\
 \tilde{\mathbf{Z}}&=\cos\alpha\, e^{\gamma} \mathbf{F}+\sin\alpha \, \star \mathbf{F},
\end{aligned}
\end{equation}
where $\alpha$ is the duality rotation angle, the field equations remain invariant, while the charges transform as follows:
\begin{equation}
    \tilde{Q}=Q\cos \alpha \quad \text{and} \quad \tilde{P}=Q\sin\alpha.
    \label{newchargs}
\end{equation}
This elegant mechanism allows the generation of dyonic configurations from purely electric solutions while preserving the spacetime geometry.

The resulting EDM BH metric maintains the familiar Schwarzschild-like form:
\begin{equation}
ds^2 = -f(r) dt^2 + \frac{dr^2}{f(r)} + r^2 (d\theta^2 + \sin^2 \theta\, d\phi^2),
\label{metric}
\end{equation}
with the modified metric lapse function:
\begin{equation}
f(r) = 1 - \frac{2M}{r} + \frac{(\tilde{Q}^2 + \tilde{P}^2) e^{-\gamma}}{r^2}.
\label{metricfunc}
\end{equation}
The exponential damping term $e^{-\gamma}$ represents the key ModMax modification, suppressing electromagnetic backreaction as $\gamma$ increases. This recovers the Reissner-Nordstr\"{o}m solution when $\gamma \to 0$.

\setlength{\tabcolsep}{12pt}     
\begin{longtable}{|c|c|c|c|}
\hline
\rowcolor{gray!50}
\textbf{$\gamma$} & \textbf{$\tilde{Q}$} & \textbf{$\tilde{P}$} & \textbf{Horizon(s)} \\
\hline
\endfirsthead

\hline
\rowcolor{gray!50}
\textbf{$\gamma$} & \textbf{$\tilde{Q}$} & \textbf{$\tilde{P}$} & \textbf{Horizon(s) [$r_h$]} \\
\hline
\endhead

0.0 & 0.0 & 0.0 & $[2.0]$ \\
\hline
0.0 & 0.5 & 0.0 & $[0.13397460,\ 1.8660254]$ \\
\hline
0.0 & 0.5 & 0.5 & $[0.29289322,\ 1.7071068]$ \\
\hline
\rowcolor{red!40}
0.0 & 1.0 & 1.0 & $[\,]$ \\
\hline
0.5 & 0.5 & 0.0 & $[0.078931417,\ 1.9210686]$ \\
\hline
0.5 & 0.5 & 1.0 & $[0.50823108,\ 1.4917689]$ \\
\hline
0.5 & 1.0 & 0.0 & $[0.37272866,\ 1.6272713]$ \\
\hline
\rowcolor{red!40}
0.5 & 1.0 & 1.0 & $[\,]$ \\
\hline
1.0 & 0.5 & 0.0 & $[1.9529062]$ \\
\hline
1.0 & 0.5 & 0.5 & $[0.096639452,\ 1.9033605]$ \\
\hline
1.0 & 1.0 & 0.5 & $[0.26505055,\ 1.7349495]$ \\
\hline
1.0 & 1.0 & 1.0 & $[0.48595611,\ 1.5140439]$ \\
\hline

\caption{\footnotesize 
Horizons obtained for selected values of the ModMax nonlinearity parameter $\gamma$, electric charge $\tilde{Q}$, and magnetic charge $\tilde{P}$ for the dyonic ModMax BH spacetime. The geometry is static, spherically symmetric, and asymptotically flat. The case $\gamma = 0$, $\tilde{Q} = 0$, $\tilde{P} = 0$ corresponds to the Schwarzschild BH. Configurations with two real roots represent \textit{non-extremal} BHs, single-root cases correspond to \textit{extremal} BHs, and empty brackets $"[\,]"$ denote \textit{naked} BHs.
}
\label{tab:dyonic_modmax}
\end{longtable}
\setlength{\tabcolsep}{10pt}

\begin{figure}[H]
    \centering
    \begin{subfigure}{0.31\textwidth}
        \centering
        \includegraphics[width=\textwidth]{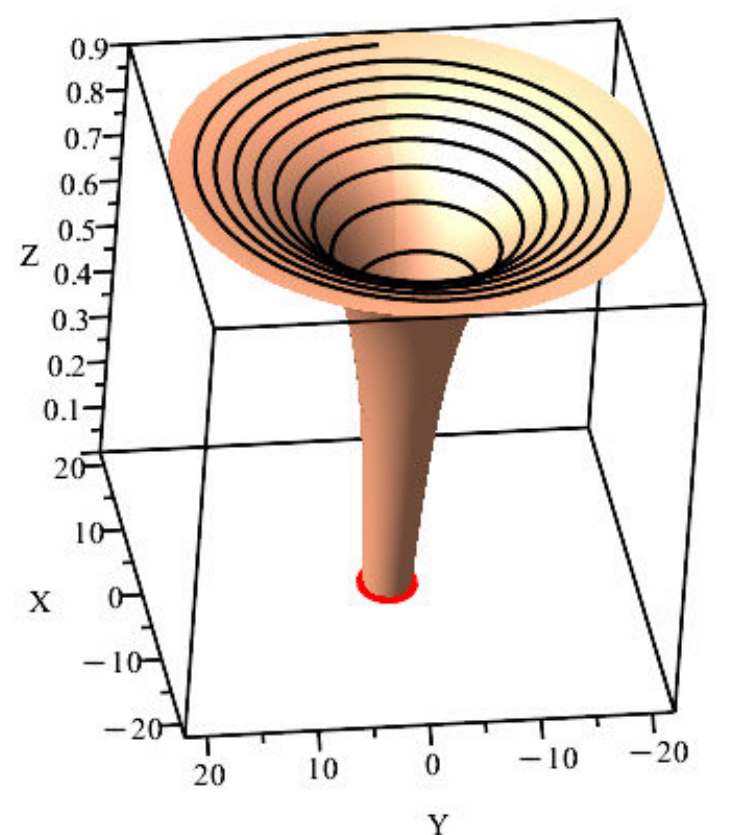}
        \caption{[$\gamma=0$, $\tilde{Q}=0$, $\tilde{P}=0$] Schwarzschild BH.}
        \label{fig:dyon-a}
    \end{subfigure}
    \hfill
    \begin{subfigure}{0.31\textwidth}
        \centering
        \includegraphics[width=\textwidth]{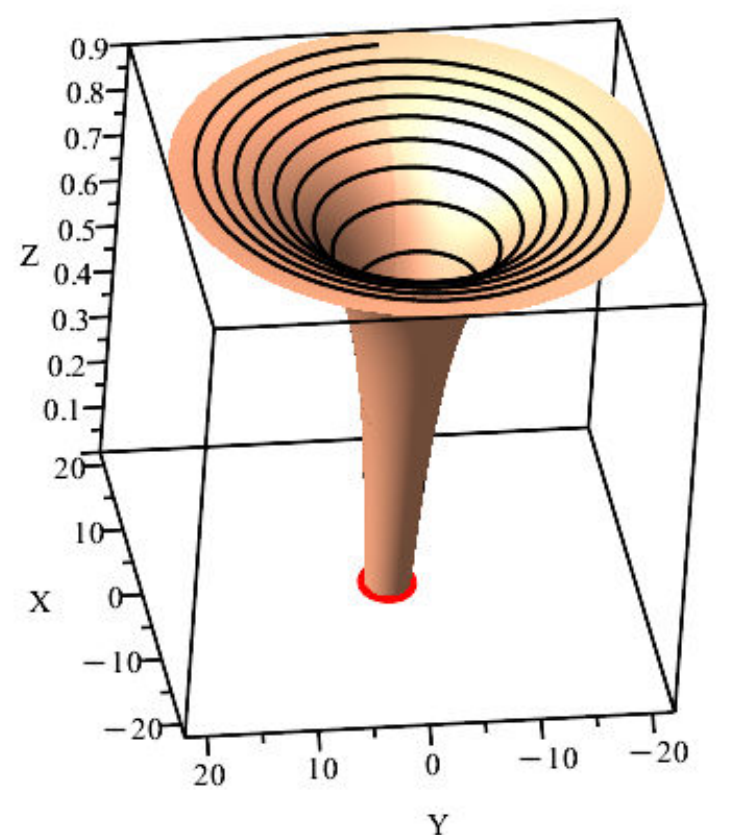}
        \caption{[$\gamma=0.5$, $\tilde{Q}=0.5$, $\tilde{P}=1$] Dyonic ModMax BH.}
        \label{fig:dyon-b}
    \end{subfigure}
    \hfill
    \begin{subfigure}{0.31\textwidth}
        \centering
        \includegraphics[width=\textwidth]{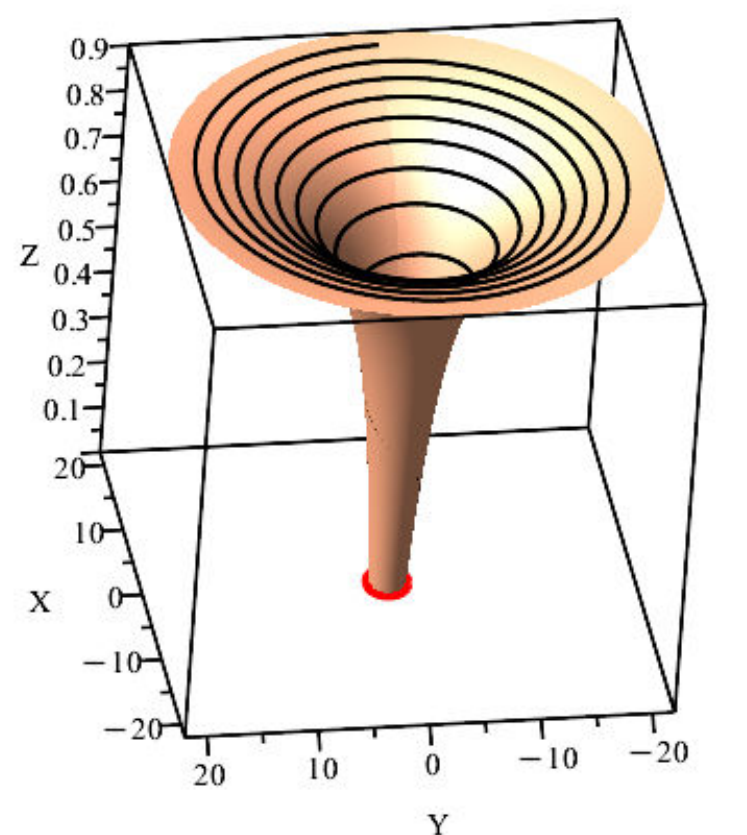}
        \caption{[$\gamma=1$, $\tilde{Q}=1$, $\tilde{P}=1$] Dyonic ModMax BH.}
        \label{fig:dyon-c}
    \end{subfigure}
    \caption{Embedding diagrams for Schwarzschild and selected dyonic ModMax BH configurations with varying ModMax nonlinearity $\gamma$, electric charge $\tilde{Q}$, and magnetic charge $\tilde{P}$. The mass is fixed to $M=1$. These plots illustrate how increasing the charge and nonlinearity affects the geometry and horizon structure. Parameter choices correspond to horizon configurations listed in Table~\ref{tab:dyonic_modmax}. The red rings denote the event horizons.}
    \label{fig:dyonic_embeddings}
\end{figure}

Table~\ref{tab:dyonic_modmax} demonstrates how the horizon structure depends on the ModMax parameter $\gamma$ and charges $\tilde{Q}$, $\tilde{P}$. Non-extremal BHs exhibit two horizons, extremal cases show single horizons, while naked singularities (highlighted in red) emerge for certain parameter combinations. The embedding diagrams in Figure~\ref{fig:dyonic_embeddings} visualize these geometric modifications, clearly showing how increasing charges and nonlinearity affect the spacetime curvature and horizon formation. The red rings in the figures mark the event horizons, providing intuitive visualization of how the ModMax parameter $\gamma$ modifies the BH geometry compared to the classical Schwarzschild case \cite{frolov2012black}.

\section{Hawking Radiation of Dyonic ModMax BHs via Quantum Tunneling} \label{iz3}

To explore the thermal spectrum of BHs within the EDM framework, we analyze the tunneling of scalar particles using the Hamilton-Jacobi technique \cite{Gursel:2015hka}. In this semiclassical approach, quantum fluctuations near the event horizon allow particles to escape via classically forbidden trajectories, and the tunneling amplitude relates to the imaginary part of the particle's action.

We begin with a minimally coupled scalar field $\Phi$ of mass $m_p$ propagating in the background geometry of a charged dyonic BH described by the ModMax-modified lapse function $f(r)$:
\begin{equation}
\square \Phi - m_p^2 \Phi = 0, \label{kg_eqn}
\end{equation}
where the box operator is defined with respect to the curved spacetime metric. Assuming spherical symmetry and a time-radial ansatz, the field equation reduces to:
\begin{equation}
-\frac{\partial^2 \Phi}{\partial t^2} + f(r)\frac{\partial}{\partial r} \left(f(r) \frac{\partial \Phi}{\partial r} \right) - m_p^2 f(r)^2 \Phi = 0.
\end{equation}
Applying the WKB approximation \cite{Sakalli:2022xrb}, we express the scalar field in terms of a rapidly varying action $I(t, r)$:
\begin{equation}
\Phi(t, r) = \exp\left(-\frac{i}{\hbar} I(t, r)\right).
\end{equation}
Substituting this ansatz into the Klein-Gordon equation and considering the leading-order contribution in $\hbar$ gives the Hamilton-Jacobi equation:
\begin{equation}
-(\partial_t I)^2 + f(r)^2 (\partial_r I)^2 + m_p^2 f(r)^2 = 0.
\end{equation}
To proceed, we assume a separable solution of the action in the form:
\begin{equation}
I(t, r) = -\omega t + W(r),
\end{equation}
where $\omega$ denotes the energy of the emitted particle. Inserting this into the previous equation yields:
\begin{equation}
(W')^2 = \frac{\omega^2 - m_p^2 f(r)^2}{f(r)^4}.
\end{equation}
To isolate the behavior near the BH event horizon $r = r_h$, where $f(r_h) = 0$, we expand the metric function linearly as:
\begin{equation}
f(r) \approx f'(r_h)(r - r_h) + \cdots. \label{f_expansion}
\end{equation}
This leads to a divergent integral for the radial action as follows:
\begin{equation}
W(r) = \pm \int \frac{\omega}{f'(r_h)} \frac{dr}{r - r_h}.
\end{equation}
The integral contains a simple pole at the horizon and is evaluated via contour integration in the complex plane:
\begin{equation}
\text{Im}(W) = \frac{\pi \omega}{f'(r_h)}.
\end{equation}
The probability of tunneling across the horizon is then proportional to the exponential of the imaginary part of the action according to:
\begin{equation}
\Gamma \sim \exp\left(-\frac{2\pi \omega}{f'(r_h)}\right).
\end{equation}
Comparing this result with the Boltzmann factor $\exp(-\omega/T)$, the present Hawking temperature turns out to be as:
\begin{equation}
T_H = \frac{f'(r_h)}{2\pi}.
\end{equation}
Now, consider the metric lapse function $f(r)$ in the dyonic ModMax solution as follows:
\begin{equation}
f(r) = 1 - \frac{2M}{r} + \frac{(\tilde{Q}^2 + \tilde{P}^2) e^{-\gamma}}{r^2}.
\end{equation}
Thereafter, taking the derivative at the horizon radius $r_h$ gives:
\begin{equation}
f'(r_h) = \frac{2M}{r_h^2} - \frac{2(\tilde{Q}^2 + \tilde{P}^2) e^{-\gamma}}{r_h^3}.\label{f'}
\end{equation}
Substituting into the temperature formula yields:
\begin{equation}
T_H = \frac{1}{2\pi} \left( \frac{M}{r_h^2} - \frac{(\tilde{Q}^2 + \tilde{P}^2)e^{-\gamma}}{r_h^3} \right). \label{hawking_modmax}
\end{equation}

\begin{figure}[ht!]
    \centering
    \includegraphics[width=0.55\textwidth]{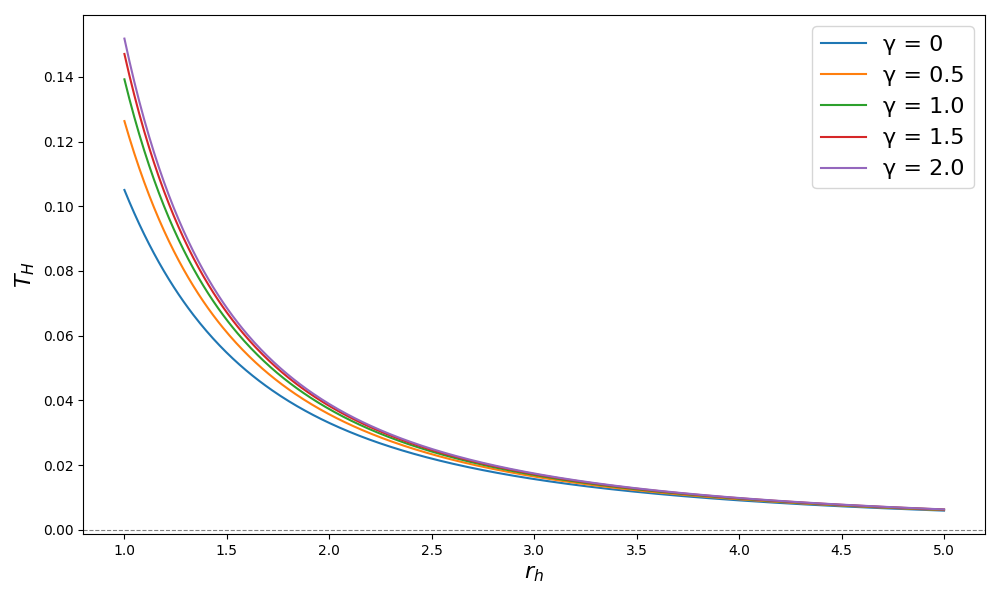}
    \caption{Hawking temperature $T_H$ as a function of horizon radius $r_h$ for various values of the ModMax parameter $\gamma$. Parameters: $M = 1.0$, $\tilde{Q} = 0.5$, $\tilde{P} = 0.3$. The charge contribution to $T_H$ is multiplied by $e^{-\gamma}$, so increasing $\gamma$ suppresses the electromagnetic correction and drives $T_H$ toward the Schwarzschild value. All curves converge to the same envelope at large $r_h$ as $\gamma$ grows.}
    \label{hawking}
\end{figure}

This result demonstrates the thermal behavior of EDM BHs and reveals that both electric and magnetic charges reduce the temperature, an effect modulated by the nonlinear parameter $\gamma$. The exponential damping factor $e^{-\gamma}$ significantly alters the charge contribution to the temperature, providing a smooth interpolation between NLE and standard Einstein-Maxwell theory. In the limit $\gamma \rightarrow 0$, the standard Reissner-Nordstr\"{o}m-like behavior is recovered.

Figure~\ref{hawking} illustrates the temperature profile as a function of the horizon radius for different values of $\gamma$. The curves show that increasing $\gamma$ reduces the influence of electromagnetic charges on the thermal emission, effectively stabilizing the temperature at larger horizon radii. This behavior has profound implications for BH evaporation dynamics and the formation of stable remnants in the final stages of Hawking radiation \cite{nozari2008hawking}.

\section{Quantum Tunneling of Massive Scalar Fields with GUP Corrections in EDM BH Geometry} \label{iz4}

In this section, we explore the effects of quantum gravity on the Hawking radiation of massive scalar particles emitted from a static, charged BH, whose geometry is described by the line element Eq.\eqref{metric}. To incorporate quantum gravity effects, we utilize a Klein-Gordon framework corrected by GUP \cite{Todorinov:2020jtq}. The modified equation for a scalar field $\Phi$ to first order in $\beta$ becomes:
\begin{equation}
- \hbar^2 \partial_t^2 \Phi = \left[ -\hbar^2 \nabla^2 + m^2 \right] \left[1 - 2\beta \left( -\hbar^2 \nabla^2 + m^2 \right) \right] \Phi,
\end{equation}
where $m$ is the mass of the scalar field, and $\beta$ is the GUP parameter encoding minimal length effects. The GUP parameter $\beta$ is commonly expressed as $\beta = \beta_{0}\,\ell_{p}^{2}/\hbar^{2}$, where $\ell_{p}$ denotes the Planck length and $\beta_{0}$ is a dimensionless constant encoding the strength of quantum gravity effects. Within this framework, the tunneling formalism remains valid in the semiclassical regime while consistently incorporating leading-order minimal-length corrections. Adopting a semiclassical approximation, we take the ansatz:
\begin{equation}
\Phi(t, r, \theta, \phi) = \exp\left[ \frac{i}{\hbar} I(t, r, \theta, \phi) \right],
\end{equation}
and expand the action $I$ as:
\begin{equation}
I(t, r, \theta, \phi) = -Et + R(r) + j \phi.
\end{equation}
Assuming the emission occurs at a fixed polar angle $\theta = \theta_0$, and keeping leading order contributions, the radial equation simplifies to a quartic form in $\partial_r R$. Upon solving, the radial action integral becomes:
\begin{equation}
R(r) = \pm \int \frac{ \sqrt{E^2 - f(r)\left[ m^2 + \frac{j^2}{r^2 \sin^2\theta} - 2\beta \lambda \right]} }{ f(r) \left[ 1 + \beta \left( m^2 + \frac{E^2}{f(r)} + \frac{j^2}{r^2 \sin^2\theta} \right) \right] }\, dr,
\end{equation}
in which $\lambda$ is:
\begin{equation}
    \lambda=\frac{j^4}{r^4 \sin^4\theta} + \frac{2m^2 j^2}{r^2 \sin^2\theta} + m^4. 
\end{equation}
Close to the event horizon $r = r_h$, where $f(r_h) = 0$, we expand the integrand to isolate the pole contribution. The resulting imaginary part of the action reads:
\begin{equation}
\operatorname{Im} R(r_h) = \frac{\pi E r_h^2}{f'(r_h)}\left(1 + \beta \xi \right),
\end{equation}
with the GUP correction term given explicitly as:
\begin{equation}
\xi = \frac{m^2}{2} + \frac{j^2}{2 r_h^2 \sin^2\theta}.
\end{equation}
To resolve the well-known "factor of two problem," we employ a contour integral in momentum space under canonical transformations and include both spatial and temporal contributions \cite{nozari2013tunneling}. This gives the complete tunneling probability:
\begin{equation}
\Gamma \sim \exp\left[-\frac{4\pi E r_h^2}{f'(r_h)}(1 + \beta \zeta)\right].
\end{equation}
The derivative of the metric function evaluated at the horizon is computed in Eq.\eqref{f'}. Using the relation between tunneling rate and temperature via the Boltzmann factor, $\Gamma \sim e^{-E/T_{GUP}}$, the effective temperature observed by an asymptotic observer is:
\begin{equation}
T_{GUP} = \frac{f'(r_h)}{4\pi r_h^2} \left(1 - \beta \zeta \right),
\end{equation}
which explicitly gives:
\begin{equation}
T_{GUP} = \frac{1}{4\pi} \left( \frac{2M}{r_h^2} - \frac{2(\tilde{Q}^2 + \tilde{P}^2) e^{-\gamma}}{r_h^3} \right) \left(1 - \frac{\beta}{2} \left[m^2 + \frac{j^2}{r_h^2 \sin^2\theta} \right] \right).
\end{equation}

This expression reveals that the GUP-corrected Hawking temperature depends not only on the BH parameters and charges, but also sensitively on the energy, mass, and angular momentum of the emitted scalar particle \cite{gecim2017gup}. The presence of the GUP correction term effectively slows down the increase in temperature during evaporation, implying the potential formation of a stable remnant in the late stage of BH decay.

The GUP modifications introduce several physically significant effects. First, the correction term $\beta \zeta$ depends on the particle's mass and angular momentum, making the radiation spectrum particle-dependent rather than universal. Second, the negative sign in the correction factor $(1 - \beta \zeta)$ indicates that GUP effects reduce the effective temperature, leading to slower evaporation rates compared to classical predictions. This suppression becomes more pronounced for massive particles and those with large angular momentum. The implications for BH thermodynamics are profound. As the BH shrinks through Hawking radiation, the GUP corrections become increasingly important, eventually preventing the temperature from diverging. This mechanism naturally leads to the formation of Planck-scale remnants, providing a potential resolution to the information paradox and offering candidates for dark matter \cite{adler2001generalized,chen2015black}.

\section{Deflection Angle in Plasma Environment via GBT} \label{iz5}

The spacetime geometry that describes the EDM BH, which is static and spherically symmetric, is characterized by the line element Eq.\eqref{metric}. To explore photon trajectories (null geodesics), we confine our analysis to the equatorial plane ($\theta = \pi/2$), simplifying the condition $ds = 0$ to yield:
\begin{equation}
dt^2 = \frac{dr^2}{f(r)^2} + \frac{r^2 d\phi^2}{f(r)}.
\end{equation}
Considering light propagation in a dispersive medium, specifically a plasma, we introduce a refractive index \cite{crisnejo2018weak}:
\begin{equation}
n(r) = \sqrt{1 - \frac{\omega_e^2}{\omega_\infty^2}} f(r),
\end{equation}
where $\omega_e$ and $\omega_\infty$ represent the plasma frequency and the photon frequency measured at infinity, respectively. In this work, we assume a homogeneous plasma distribution, for which the plasma frequency $\omega_e$ is taken to be spatially constant. This choice allows an analytic treatment within the Gauss--Bonnet framework and captures the leading-order dispersive effects of the plasma on photon trajectories. The resulting effective optical metric is then:
\begin{equation}
d\sigma^2 = n^2(r) \left( \frac{dr^2}{f(r)} + r^2 d\phi^2 \right).
\end{equation}
Next, the Gaussian curvature associated with this optical geometry is computed from half the Ricci scalar:
\begin{equation}
K = \frac{1}{2} \text{RicciScalar}.
\end{equation}
Expanding this expression to leading order yields the following:
\begin{align}
K \approx{} & \frac{3 M^{2}}{r^{4}}-\frac{2 M}{r^{3}}-\frac{3 M \delta}{r^{3}}+\frac{12 M^{2} \delta}{r^{4}}-\frac{12 M^{3} \delta}{r^{5}}+\frac{3 \tilde{P}^{2}}{r^{4} {\mathrm e}^{\gamma}}+\frac{3 \tilde{Q}^{2}}{r^{4} {\mathrm e}^{\gamma}}+\frac{5 \tilde{P}^{2} \delta}{r^{4} {\mathrm e}^{\gamma}}+\frac{5 \tilde{Q}^{2} \delta}{r^{4} {\mathrm e}^{\gamma}} \notag \\
& -\frac{6 M \tilde{P}^{2}}{r^{5} {\mathrm e}^{\gamma}}-\frac{6 M \tilde{Q}^{2}}{r^{5} {\mathrm e}^{\gamma}}-\frac{26 M \tilde{P}^{2}\delta}{r^{5}  {\mathrm e}^{\gamma}}-\frac{26 M \tilde{Q}^{2} \delta}{r^{5}  {\mathrm e}^{\gamma}}+\frac{32 M^{2} \tilde{P}^{2} \delta}{r^{6}  {\mathrm e}^{\gamma}}+\frac{32 M^{2} \tilde{Q}^{2} \delta}{r^{6}  {\mathrm e}^{\gamma}}+\frac{2 \tilde{P}^{4}}{r^{6} \left({\mathrm e}^{\gamma}\right)^{2}}+\frac{2 \tilde{Q}^{4}}{r^{6} \left({\mathrm e}^{\gamma}\right)^{2}}+\frac{4 \tilde{P}^{2} \tilde{Q}^{2}}{r^{6} \left({\mathrm e}^{\gamma}\right)^{2}} \notag \\
& +\frac{10 \tilde{P}^{4} \delta}{r^{6}  \left({\mathrm e}^{\gamma}\right)^{2}}+\frac{10 \tilde{Q}^{4} \delta}{r^{6}  \left({\mathrm e}^{\gamma}\right)^{2}}+\frac{20 \tilde{P}^{2} \tilde{Q}^{2} \delta}{r^{6}  \left({\mathrm e}^{\gamma}\right)^{2}}-\frac{23 M \tilde{P}^{4} \delta}{r^{7} \left({\mathrm e}^{\gamma}\right)^{2}}-\frac{23 M \tilde{Q}^{4} \delta}{r^{7} \left({\mathrm e}^{\gamma}\right)^{2}}-\frac{46 M \tilde{P}^{2} \tilde{Q}^{2} \delta}{r^{7} \left({\mathrm e}^{\gamma}\right)^{2}}+\frac{5 \delta \tilde{P}^{6}}{r^{8}  \left({\mathrm e}^{\gamma}\right)^{3}}+\frac{5 \delta \tilde{Q}^{6}}{r^{8}  \left({\mathrm e}^{\gamma}\right)^{3}} \notag \\
& +\frac{15 \delta \tilde{Q}^{2} \tilde{P}^{4}}{r^{8} \left({\mathrm e}^{\gamma}\right)^{3}}+\frac{15 \delta \tilde{Q}^{4} \tilde{P}^{2}}{r^{8}  \left({\mathrm e}^{\gamma}\right)^{3}},
\end{align}
where
\begin{equation}
    \delta=\frac{\omega_e^2}{\omega_{\infty}^2}.
\end{equation}

To compute the bending of light caused by the EDM BH in a plasma environment, we employed the GBT \cite{Gibbons:2008rj}. We consider a region $\mathcal{R}$ bounded by a geodesic $\gamma$ and a circular arc $C_R$, where the GBT takes the form:
\begin{equation}
\iint_{\mathcal{R}} K\, dS + \oint_{\partial \mathcal{R}} \kappa\, dt + \sum \epsilon_i = 2\pi \chi(\mathcal{R}),
\end{equation}
where $K$ is the Gaussian curvature, $\kappa$ the geodesic curvature of the boundary, and $\chi(\mathcal{R})$ is the Euler characteristic. In the weak field regime, with $R \to \infty$, this simplifies to:
\begin{equation}
\iint_{\mathcal{R}} K\, dS + \oint_{\partial \mathcal{R}} \kappa\, dt = \pi.
\end{equation}
The trajectory of the light ray in this approximation is expanded as follows \cite{Gibbons:2008rj}:
\begin{equation}
\frac{1}{r_p} = \frac{\sin \phi}{b}, 
\end{equation}
where $b$ is the impact parameter. Substituting the leading-order curvature $K$ into the integral form of the deflection angle:
\begin{equation}
\alpha = - \int_0^{\pi} \int_{r_p}^{\infty} K \sqrt{\det(g_{\text{opt}})}\, dr\, d\phi,
\end{equation}
leads to the expression for the bending angle in a uniform plasma medium:
\begin{align}
\alpha \approx{} & \frac{4 M}{b}+\frac{6 M \delta}{b}+\frac{15 M^{2} \pi}{4 b^{2}}-\frac{3 M^{2} \pi  \delta}{4 b^{2}}-\frac{3 \pi \tilde{P}^{2}}{4 b^{2} {\mathrm e}^{\gamma}}-\frac{3 \pi \tilde{Q}^{2}}{4 b^{2} {\mathrm e}^{\gamma}}-\frac{5 \pi \tilde{P}^{2} \delta}{4 b^{2} {\mathrm e}^{\gamma}}-\frac{5 \pi \tilde{Q}^{2} \delta}{4 b^{2} {\mathrm e}^{\gamma}}-\frac{4 M \tilde{P}^{2}}{3 b^{3} {\mathrm e}^{\gamma}}-\frac{4 M \tilde{Q}^{2}}{3 b^{3} {\mathrm e}^{\gamma}} \notag \\
& +\frac{44 M \tilde{P}^{2} \delta}{9 b^{3} {\mathrm e}^{\gamma}}+\frac{44 M \tilde{Q}^{2} \delta}{9 b^{3} {\mathrm e}^{\gamma}}+\frac{27 M^{2} \pi \tilde{P}^{2}}{16 b^{4} {\mathrm e}^{\gamma}}+\frac{27 M^{2} \pi \tilde{Q}^{2}}{16 b^{4} {\mathrm e}^{\gamma}}+\frac{69 M^{2} \pi \tilde{P}^{2}\delta}{16 b^{4} {\mathrm e}^{\gamma}}+\frac{69 M^{2} \pi \tilde{Q}^{2} \delta}{16 b^{4} {\mathrm e}^{\gamma}}-\frac{3 \pi \tilde{P}^{4}}{16 b^{4} \left({\mathrm e}^{\gamma}\right)^{2}}-\frac{3 \pi \tilde{Q}^{4}}{16 b^{4} \left({\mathrm e}^{\gamma}\right)^{2}} \notag \\
& -\frac{3 \pi \tilde{P}^{2} \tilde{Q}^{2}}{8 b^{4} \left({\mathrm e}^{\gamma}\right)^{2}}-\frac{15 \pi \tilde{P}^{4} \delta}{16 b^{4} \left({\mathrm e}^{\gamma}\right)^{2}}-\frac{15 \pi \tilde{Q}^{4} \delta}{16 b^{4} \left({\mathrm e}^{\gamma}\right)^{2}}-\frac{15 \pi \tilde{P}^{2} \tilde{Q}^{2} \delta}{8 b^{4} \left({\mathrm e}^{\gamma}\right)^{2}}+\mathcal{O}(M^3).
\end{align}

\begin{figure}[ht!]
    \centering
    \includegraphics[width=0.55\textwidth]{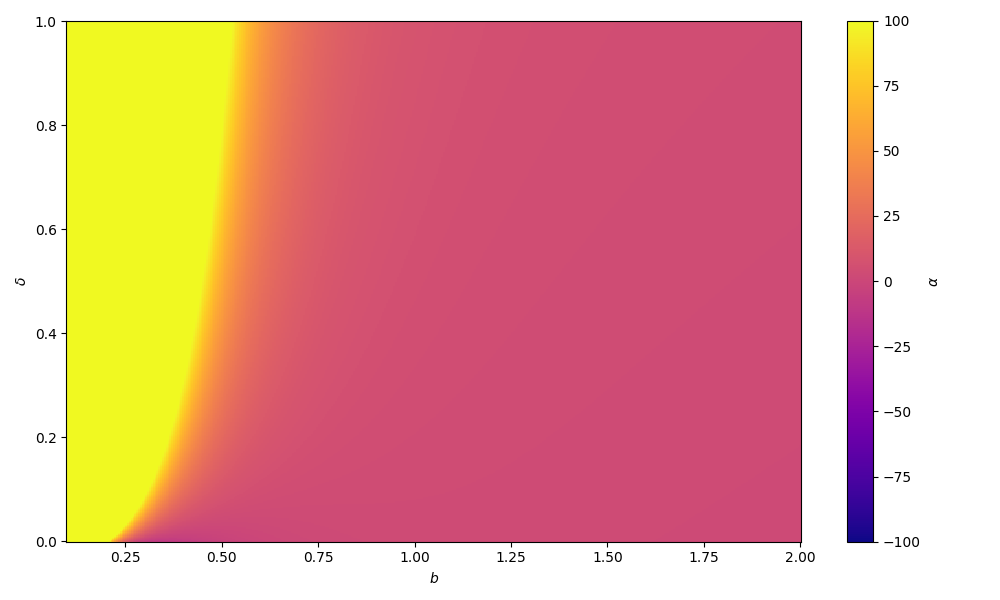}
    \caption{Deflection angle $\alpha$ of light in plasma media for parameters $M = 1.0$, $\tilde{Q} = 0.5$, $\tilde{P} = 0.5$ and $\gamma = 1$, in terms of the impact parameter $b$ and the plasma parameter $\delta$. The color scale shows the deflection angle values, with yellow regions representing large positive deflections and purple regions representing negative deflections. Strong lensing occurs at small impact parameters $b < 0.5$ and high plasma densities $\delta > 0.1$, where deflection angles approach extreme values around $100$. At large impact parameters ($b > 1.5$), the deflection approaches zero, confirming that plasma effects are localized near the BH.}
    \label{plasma1}
\end{figure}

In the limit $\delta \to 0$, the medium effect disappears, and the expression reduces to the vacuum deflection case. This agrees with previously known results up to second order in mass when $M^2$ is replaced by charge-squared terms in similar setups \cite{bisnovatyi2015gravitational}.

Figure~\ref{plasma1} visualizes the deflection angle $\alpha$ of light in a plasma medium around an EDM BH. The plot reveals that strong lensing occurs at small impact parameters and high plasma densities, where the gravitational-optical potential created by electromagnetic charges in the plasma medium generates pronounced lensing effects.

The assumption of a homogeneous plasma distribution adopted here is a simplification that permits closed-form evaluation of the GBT integrals. In astrophysical settings, the electron density near BHs typically follows a radial power-law profile $n_e(r) \propto r^{-h}$ with $h \sim 1$--$2$ for spherical accretion flows, or more complex distributions in magnetized accretion disks \cite{BisnovatyiKogan2010,Perlick2022}. When a power-law plasma frequency $\omega_p^2(r) = \kappa_0/r^\alpha$ is used instead (as we consider in Section~\ref{iz8}\,B for photon orbits), the leading-order deflection angle acquires additional $r$-dependent weighting inside the GBT surface integral, but the qualitative structure remains the same: the bending angle still increases with plasma density and decreases with the ModMax parameter $\gamma$ through the $e^{-\gamma}$ suppression factor. Quantitatively, inhomogeneous profiles redistribute the plasma correction toward the near-horizon region where $\omega_p/\omega$ is largest, typically enhancing the deflection at small impact parameters relative to the uniform case. However, the dependence on $\gamma$, $\tilde{Q}$, and $\tilde{P}$ entering through the metric function $f(r)$ is unaffected by the plasma model choice, so the ModMax-specific signatures identified here exponential damping of charge contributions and the suppressive role of $\gamma$ persist in more realistic plasma configurations. A detailed treatment with non-homogeneous plasma profiles and rotating backgrounds is left for future work.

\section{Weak Deflection Angle via the GBT Framework} \label{iz6}

To analyze the bending of light caused by EDM BHs, we restrict the motion of photons to the equatorial plane by setting $\theta = \pi/2$. This simplifies the line element into its optical counterpart, useful for studying null geodesics in the weak field regime \cite{Gibbons:2008rj,gibbons1977action}. Using this optical metric, we evaluate the Gaussian curvature $\mathcal{K}$, which approximately relates to the Ricci scalar. Up to leading terms, it reads:
\begin{equation}
\mathcal{K}(r) \approx -\frac{2 M}{r^{3}}+\frac{3 M^{2}}{r^{4}}+\frac{3 \tilde{P}^{2}}{{\mathrm e}^{\gamma} r^{4}}+\frac{3 \tilde{Q}^{2}}{{\mathrm e}^{\gamma} r^{4}}-\frac{6 M \tilde{P}^{2}}{r^{5} {\mathrm e}^{\gamma}}-\frac{6 M \tilde{Q}^{2}}{r^{5} {\mathrm e}^{\gamma}}+\frac{2 \tilde{P}^{4}}{{\mathrm e}^{2\gamma} r^{6}}+\frac{2 \tilde{Q}^{4}}{{\mathrm e}^{2\gamma} r^{6}}+\frac{4 \tilde{P}^{2} \tilde{Q}^{2}}{{\mathrm e}^{2\gamma} r^{6}}.
\end{equation}
The deflection angle $\hat{\alpha}$ is obtained by applying the GBT over a two-dimensional domain $\mathcal{D}$ that lies outside the photon trajectory \cite{Gibbons:2008rj,gibbons1977action}:
\begin{equation}
\hat{\alpha} = - \iint_{\mathcal{D}} \mathcal{K}(r)\, dS.
\end{equation}
To evaluate the surface integral, we approximate the light path as a straight line $r(\phi) = \frac{b}{\sin\phi}$, where $b$ denotes the impact parameter. The area element in polar coordinates is $dS = r\, dr\, d\phi$, yielding:
\begin{equation}
\hat{\alpha} = - \int_0^\pi \int_{\frac{b}{\sin\phi}}^\infty \mathcal{K}(r) \cdot r\, dr\, d\phi.
\end{equation}
Substituting the curvature and carrying out the integration---keeping only the most dominant contributions---the deflection angle in the weak field regime becomes:
\begin{align}
\hat{\alpha} ={} & \frac{4 M}{b}+\frac{15 M^{2} \pi}{4 b^{2}}-\frac{3 \pi \tilde{P}^{2}}{4 b^{2} {\mathrm e}^{\gamma}}-\frac{3 \pi \tilde{Q}^{2}}{4 b^{2} {\mathrm e}^{\gamma}}-\frac{4 M \tilde{P}^{2}}{3 b^{3} {\mathrm e}^{\gamma}}-\frac{4 M \tilde{Q}^{2}}{3 b^{3} {\mathrm e}^{\gamma}}+\frac{27 M^{2} \pi \tilde{P}^{2}}{16 b^{4} {\mathrm e}^{\gamma}}+\frac{27 M^{2} \pi \tilde{Q}^{2}}{16 b^{4} {\mathrm e}^{\gamma}} \notag \\
& -\frac{3 \pi \tilde{P}^{4}}{16 b^{4} {\mathrm e}^{2\gamma}}-\frac{3 \pi \tilde{Q}^{4}}{16 b^{4} {\mathrm e}^{2\gamma}}-\frac{3 \pi \tilde{P}^{2} \tilde{Q}^{2}}{8 b^{4} {\mathrm e}^{2\gamma}}-\frac{32 M \tilde{P}^{4}}{25 b^{5} {\mathrm e}^{2\gamma}}-\frac{32 M \tilde{Q}^{4}}{25 b^{5} {\mathrm e}^{2\gamma}}-\frac{64 M \tilde{P}^{2} \tilde{Q}^{2}}{25 b^{5} {\mathrm e}^{2\gamma}}+\mathcal{O}(M^3).
\end{align}

\begin{figure}[ht!]
    \centering
    \includegraphics[width=0.55\textwidth]{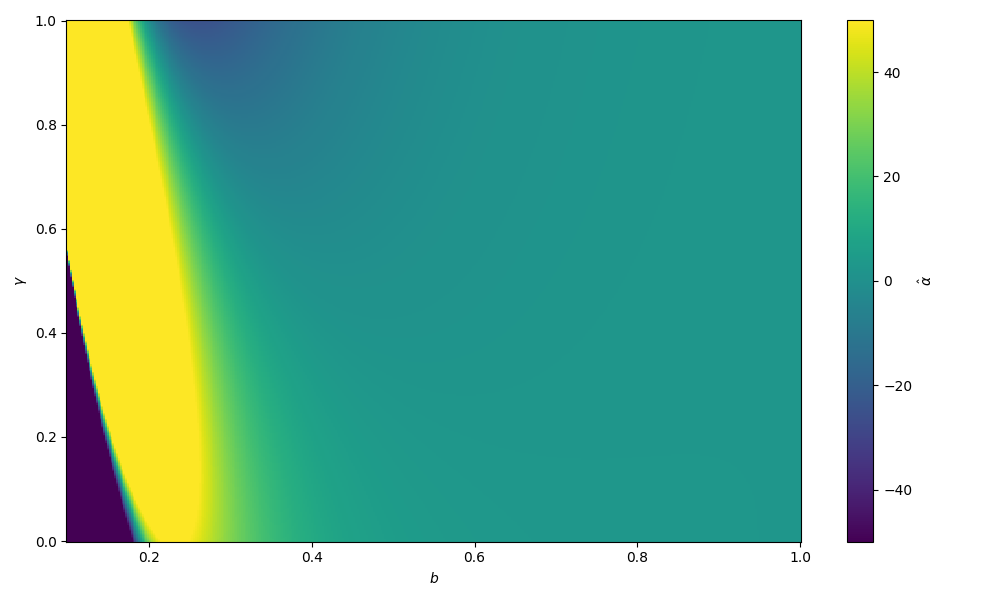}
    \caption{Deflection angle $\tilde{\alpha}$ of light in terms of the impact parameter $b$ and the ModMax parameter $\gamma$ for the case without plasma effects. Parameters: $M = 1.0$, $\tilde{Q} = 0.5$, $\tilde{P} = 0.5$. The charge-dependent terms in Eq.~(\ref{metricfunc}) carry factors of $e^{-\gamma}$ and $e^{-2\gamma}$. For $\gamma \gtrsim 1$, these terms are exponentially suppressed, and the deflection angle approaches the Schwarzschild limit $\hat{\alpha} \approx 4M/b$. In the regime $b > 0.5$ and $\gamma > 0.5$, the deflection is nearly zero beyond the pure-mass contribution.}
    \label{lensign}
\end{figure}

This expression demonstrates the rich structure of gravitational lensing in EDM theory, where the ModMax parameter $\gamma$ and dyonic charges $\tilde{Q}$, $\tilde{P}$ significantly modify the classical deflection behavior. The exponential damping factors $e^{-\gamma}$ and $e^{-2\gamma}$ appearing in various terms reflect the nonlinear electromagnetic corrections that distinguish ModMax theory from standard Einstein-Maxwell electrodynamics.

Figure~\ref{lensign} visualizes the deflection angle $\tilde{\alpha}$ in the context of EDM theory without plasma effects. The graph reveals that the deflection angle varies dramatically, particularly in regions of small $b$ and low $\gamma$ values. The visualization demonstrates that the ModMax parameter $\gamma$ exhibits a suppressive effect on the optical properties of the EDM field, while classical gravitational deflection becomes significant only for specific energy-density configurations. This framework provides crucial information for distinguishing between classical general relativity and ModMax-modified gravity through precision lensing observations, particularly in strong-field regimes where the nonlinear electromagnetic corrections become most pronounced.

\section{Deflection of Light in Axion-Plasmon Environments} \label{iz7}

Axion-photon coupling, rooted in both string theory and dark matter phenomenology, modifies electromagnetic wave propagation in magnetized plasmas \cite{buschmann2022dark,raffelt1988mixing,svrcek2006axions,marsh2016axion}. In the vicinity of an EDM BH, this coupling leads to measurable deviations in the path of photons, providing a probe of beyond-Standard-Model physics in strong gravitational fields.

We consider an effective theory incorporating gravitational, electromagnetic, and axionic components, where the axion-photon interaction introduces an additional term in the Lagrangian \cite{atamurotov2021axion}:
\begin{equation}
\mathcal{L}_{\text{int}} = -\frac{g}{4} \varepsilon^{\mu\nu\alpha\beta} F_{\mu\nu} F_{\alpha\beta}.
\end{equation}
The axion photon interaction considered here corresponds to the standard pseudoscalar coupling $g_{\phi\gamma\gamma}\,\phi\,F_{\mu\nu}\tilde{F}^{\mu\nu}$, which is widely used in axion electrodynamics \cite{Sikivie1983,raffelt1988mixing}. The plasma contribution enters through an effective refractive index, and the axion--photon coupling modifies photon propagation indirectly via dispersive effects rather than through a direct modification of the background spacetime geometry. The key physical assumption is that the axion field oscillates coherently with frequency $\omega_\varphi$ in the presence of an external magnetic field $B_0$, producing an effective mixing between photon polarization states. This mixing alters the phase velocity of electromagnetic waves in the magnetized plasma and is encoded in the modified refractive index given below. The resulting deflection angle therefore captures the combined dispersive effects of plasma, axion oscillations, and the gravitational field of the EDM BH.

The physical relevance of this model lies in the fact that strongly magnetized environments around compact objects such as magnetars ($B_0 \sim 10^{14}$--$10^{15}$\,G) or the inner regions of accretion disks around supermassive BHs ($B_0 \sim 10^{2}$--$10^{4}$\,G) are precisely the settings where axion-photon conversion is expected to be most efficient \cite{Hook2018}.

In such a background, photons acquire an effective refractive index depending on the plasma frequency $\omega_p$, the axion frequency $\omega_\varphi$, and the external magnetic field $B_0$. A simplified expression for the refractive index is \cite{atamurotov2021axion}:
\begin{equation}
n(r) \simeq \sqrt{1 - \frac{\omega_p^2}{\omega_0^2} f(r) \left(1 + \frac{B_0^2}{1 - \omega_\varphi^2} \right)},
\end{equation}
where $f(r)$ is the BH metric function and $\omega_0$ is a reference frequency. This modified refractive index significantly alters the optical geometry felt by light rays, incorporating both plasma dispersion effects and axion-induced birefringence.

By applying the GBT to this effective optical manifold, the deflection angle of light can be expressed as:
\begin{align}
\Theta \simeq{} & \frac{1}{B_{0}^{2}-\omega_{\varphi}^{2}+1}\Bigg[-\frac{2 \sigma M}{b}+\frac{2 \sigma M \omega_{\varphi}^{2}}{b}+\frac{\sigma \pi \tilde{P}^{2}}{4 {\mathrm e}^{\gamma} b^{2}}+\frac{\sigma \pi \tilde{Q}^{2}}{4 {\mathrm e}^{\gamma} b^{2}}+\frac{\sigma \pi \tilde{P}^{2} \omega_{\varphi}^{2}}{4 {\mathrm e}^{\gamma} b^{2}}-\frac{\sigma \pi \tilde{Q}^{2} \omega_{\varphi}^{2}}{4 {\mathrm e}^{\gamma} b^{2}} \notag \\
& +\frac{4 \sigma M \tilde{P}^{2}}{3 {\mathrm e}^{\gamma} b^{3}}+\frac{4 \sigma M \tilde{Q}^{2}}{3 {\mathrm e}^{\gamma} b^{3}}-\frac{4 \sigma M \tilde{P}^{2} \omega_{\varphi}^{2}}{3 {\mathrm e}^{\gamma} b^{3}}-\frac{4 \sigma M \tilde{Q}^{2} \omega_{\varphi}^{2}}{3 {\mathrm e}^{\gamma} b^{3}}\Bigg]+\mathcal{O}(M^2),
\end{align}
where $b$ is the impact parameter, and $\sigma=\frac{\omega_{0}^{2}}{\omega_{p}^{2}}$.

\begin{figure}[ht!]
    \centering
    \includegraphics[width=0.55\textwidth]{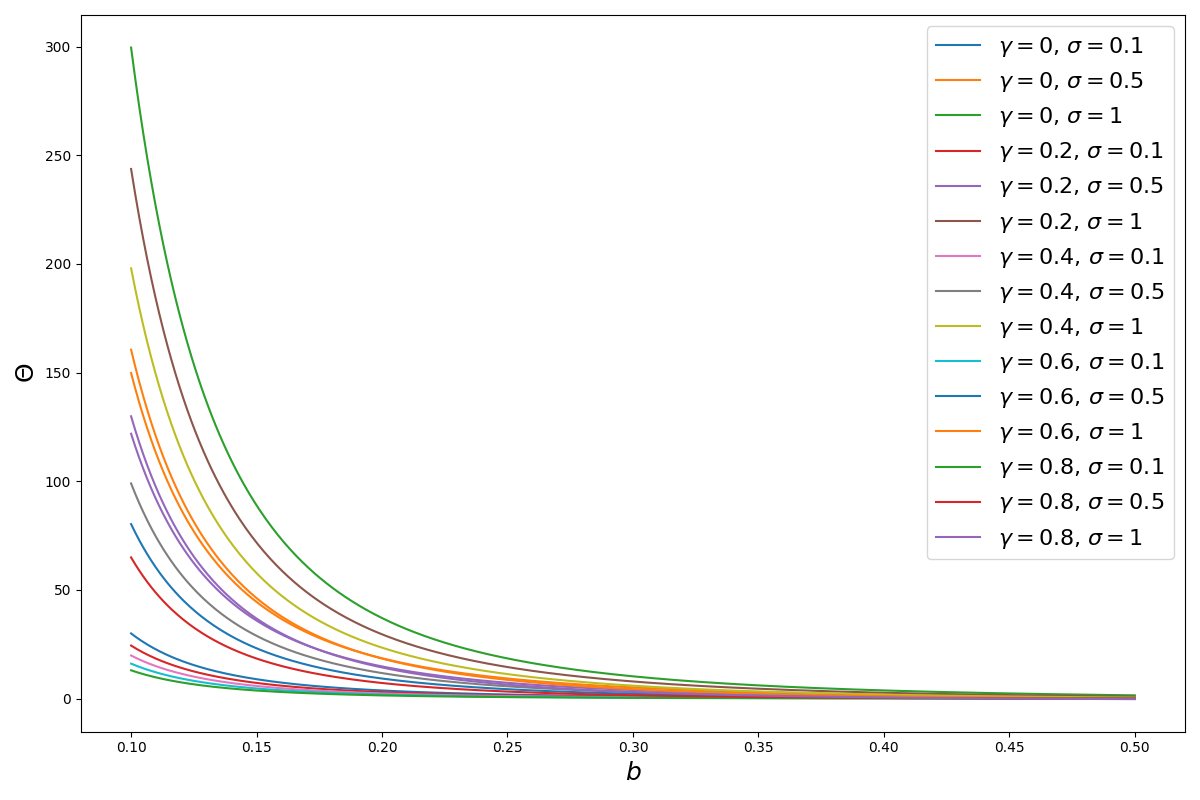}
    \caption{Deflection angle $\Theta$ in axion-plasmon environments as a function of the impact parameter $b$, plotted for a grid of ModMax parameter values $\gamma \in \{0,\,0.2,\,0.4,\,0.6,\,0.8\}$ and ratio $\sigma=\omega_{0}^{2}/\omega_{p}^{2} \in \{0.1,\,0.5,\,1\}$. Remaining parameters are fixed at $M = 1.0$, $\tilde{Q} = \tilde{P} = 0.5$, $\omega_\varphi = 0.5$, and $B_0 = 1.0$. The family of curves illustrates how the deflection decreases rapidly with $b$ and how increasing either $\gamma$ (stronger exponential charge damping) or decreasing $\sigma$ (relatively stronger plasma response) reduces the axion-plasmon-mediated bending contribution.}
    \label{sss}
\end{figure}

\begin{figure}[ht!]
    \centering
    \includegraphics[width=0.55\textwidth]{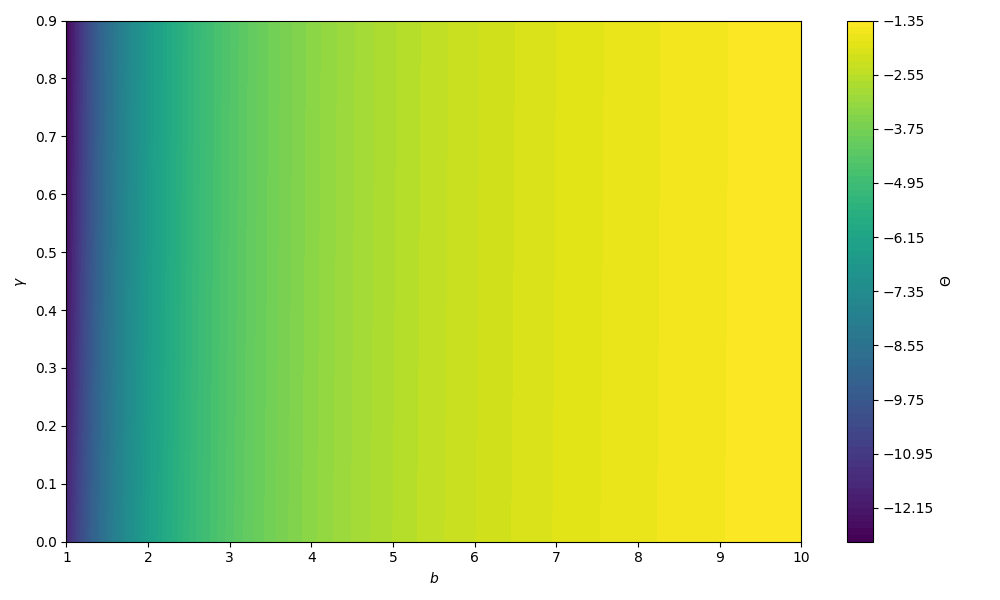}
    \caption{Density plot of deflection angle $\Theta$ as a function of impact parameter $b$ and ModMax parameter $\gamma$ in axion-plasmon environments. Fixed parameters: $M = 1.0$, $\tilde{Q} = \tilde{P} = 0.5$, $\omega_0 = 1.0$, $\omega_\varphi = 0.5$, $\omega_p = 0.5$, and $B_0 = 1.0$. The color gradients reveal regions where axion effects either enhance or suppress the deflection, depending on the relative magnitudes of the axion frequency, magnetic field strength, and plasma density.}
    \label{lensign_axion}
\end{figure}

This framework reveals that axions induce distinctive, frequency-dependent modifications to the gravitational lensing signal \cite{atamurotov2021axion}. These signatures become especially pronounced near resonance with the axion mass scale, offering a novel astrophysical channel to probe axionic dark matter in strongly gravitating, magnetized environments.

Figure~\ref{sss} plots the deflection angle $\Theta$ against the impact parameter $b$ for a representative grid of $\gamma$ and $\sigma$ values, while Figure~\ref{lensign_axion} presents a detailed density plot of the deflection angle $\Theta$ as a function of the impact parameter $b$ and ModMax parameter $\gamma$.

The denominator factor $(B_0^2 - \omega_\varphi^2 + 1)$ in the deflection formula indicates potential resonant behavior when $B_0^2 \approx \omega_\varphi^2 - 1$, suggesting that certain magnetic field configurations could dramatically amplify the lensing signal \cite{bisnovatyi2015gravitational,rogers2015frequency}. This resonant enhancement provides a powerful diagnostic tool for detecting axion dark matter through gravitational lensing observations of EDM BHs in astrophysical environments with strong magnetic fields, such as those found near magnetars or in the vicinity of supermassive BHs with accretion disks \cite{Hook2018}.

\section{Photon Motion in ModMax BHs} \label{iz8}

We explore photon dynamics in the spacetime of a static, spherically symmetric ModMax BH, described by the metric in Eq.~\eqref{metric}. To study light propagation in a plasma medium, we consider the refractive properties of the medium \cite{synge1960relativity}. The Hamiltonian for a photon in this environment, accounting for plasma effects, is:
\begin{equation}
\mathcal{H}(x^\mu, p_\mu) = \frac{1}{2} \left[ g^{\mu\nu} p_\mu p_\nu + (\omega_p^2 - 1) (p_\mu u^\mu)^2 \right],
\end{equation}
where $x^\mu$ is the photon position, $p_\mu$ its four-momentum, and $u^\mu$ the plasma four-velocity. The plasma frequency $\omega_p$ depends on local plasma conditions.

The refractive index $n$, the ratio of phase velocity to the vacuum speed of light, is \cite{mendoncca2020axion,crisnejo2018weak,sucu2025exploring}:
\begin{equation}
n^2 = 1 - \frac{\omega_p^2}{\omega^2}.
\end{equation}
The photon frequency $\omega(r)$, observed from infinity, shifts with radial position due to gravity:
\begin{equation}
\omega(r) = \frac{\omega_0}{\sqrt{f(r)}},
\end{equation}
where $\omega_0$ is the photon's energy at infinity and $f(r)$ the metric function. In sparse plasmas where $\omega_p \ll \omega$, plasma effects diminish, and light behaves as in a vacuum. Setting $\omega_p = 0$ recovers the vacuum case, where $n = 1$.

Photon trajectories follow geodesic equations. In the equatorial plane ($\theta = \pi/2$, $p_\theta = 0$), the four-velocity components are:
\begin{equation}
\dot{t} = \frac{-p_t}{f(r)}, \quad \dot{r} = p_r f(r), \quad \dot{\varphi} = \frac{p_\varphi}{r^2},
\end{equation}
where the dot denotes differentiation with respect to the affine parameter $\lambda$. From the Hamiltonian condition $\mathcal{H} = 0$, the radial motion is:
\begin{equation}
\frac{dr}{d\varphi} = \frac{r^2 f(r) p_r}{p_\varphi}.
\end{equation}
Using $p_r$ and $p_\varphi$, the path of the photon satisfies the following conditions:
\begin{equation}
\frac{dr}{d\varphi} = r \sqrt{f(r)} \sqrt{k^2(r) \frac{\omega_0}{p_\varphi} - 1},
\end{equation}
where $k^2(r)$ is \cite{sucu2025exploring}:
\begin{equation}
k^2(r) = r^2 \left( \frac{1}{f(r)} - \frac{\omega_p^2(r)}{\omega_0^2} \right).
\end{equation}

The radius of the photon sphere $r_p$, where the photons orbit circularly, is found by solving the following:
\begin{equation}
\frac{d}{dr} \left( r^2 \left[ \frac{1}{f(r)} - \frac{\omega_p^2(r)}{\omega_0^2} \right] \right) \Bigg|_{r=r_p} = 0.
\end{equation}
For specific plasma models (homogeneous or power-law), we solve this numerically to find $r_p$. The condition becomes:
\begin{equation}
\frac{2}{f(r_p)} - \frac{2 \omega_p^2(r_p)}{\omega_0^2} - r_p \left( \frac{f'(r_p)}{f(r_p)^2} + \frac{2 \omega_p(r_p) \omega_p'(r_p)}{\omega_0^2} \right) = 0, \label{eq:sphere_condition}
\end{equation}
with $f'(r_p)$ as the radial derivative of $f(r)$.

\subsection{Constant Plasma Density}

For a homogeneous plasma with fixed $\omega_p$, Eq.~\eqref{eq:sphere_condition} reduces to:
\begin{equation}
\frac{2}{f(r_p)} - \frac{2 \omega_p^2}{\omega_0^2} = r_p \frac{f'(r_p)}{f(r_p)^2}.
\end{equation}

\begin{table}[ht!]
\centering
\begin{tabular}{|c|c|c|c|c|}
\hline
\(\tilde{Q}\) & \(\tilde{P}\) & \(\gamma\) & \(\omega_0\) & \(\omega_p\) \\
\hline
0 & 0.5 & 0 & 0.5 & 0.3264 \\
0 & 0.5 & 0 & 1   & 0.6527 \\
0 & 0.5 & 1 & 0.5 & 0.2081 \\
0 & 0.5 & 1 & 1   & 0.4161 \\
0 & 1   & 0 & 0.5 & 0.5303 \\
0 & 1   & 0 & 1   & 1.0607 \\
0 & 1   & 1 & 0.5 & 0.3820 \\
0 & 1   & 1 & 1   & 0.7641 \\
0.5 & 0 & 0 & 0.5 & 0.3264 \\
0.5 & 0 & 0 & 1   & 0.6527 \\
0.5 & 0 & 1 & 0.5 & 0.2081 \\
0.5 & 0 & 1 & 1   & 0.4161 \\
0.5 & 0.5 & 0 & 0.5 & 0.4286 \\
0.5 & 0.5 & 0 & 1   & 0.8571 \\
0.5 & 0.5 & 1 & 0.5 & 0.2857 \\
0.5 & 0.5 & 1 & 1   & 0.5715 \\
0.5 & 1 & 0 & 0.5 & 0.5580 \\
0.5 & 1 & 0 & 1   & 1.1161 \\
0.5 & 1 & 1 & 0.5 & 0.4158 \\
0.5 & 1 & 1 & 1   & 0.8315 \\
1 & 0 & 0 & 0.5 & 0.5303 \\
1 & 0 & 0 & 1   & 1.0607 \\
1 & 0 & 1 & 0.5 & 0.3820 \\
1 & 0 & 1 & 1   & 0.7641 \\
1 & 0.5 & 0 & 0.5 & 0.5580 \\
1 & 0.5 & 0 & 1   & 1.1161 \\
1 & 0.5 & 1 & 0.5 & 0.4158 \\
1 & 0.5 & 1 & 1   & 0.8315 \\
1 & 1 & 0 & 0.5 & 0.6000 \\
1 & 1 & 0 & 1   & 1.2000 \\
1 & 1 & 1 & 0.5 & 0.4871 \\
1 & 1 & 1 & 1   & 0.9741 \\
\hline
\end{tabular}
\caption{Variation of the plasma frequency $\omega_p$ for different values of electric charge $\tilde{Q}$, magnetic charge $\tilde{P}$, nonlinearity parameter $\gamma$, and photon energy at infinity $\omega_0$, assuming a constant plasma density. Increasing $\gamma$ reduces $\omega_p$, reflecting the weakened electromagnetic confinement caused by the $e^{-\gamma}$ suppression. The system reduces to the classical Reissner-Nordstr\"{o}m limit when $\gamma \to 0$.}\label{homojen}
\end{table}

Table~\ref{homojen} illustrates how the plasma frequency $\omega_p$ varies with different values of electric charge $\tilde{Q}$, magnetic charge $\tilde{P}$, damping parameter $\gamma$, and photon energy at infinity $\omega_0$, assuming a constant plasma density. The exponential damping term $e^{-\gamma}$ captures the nonlinear correction, significantly reducing the backreaction of the electromagnetic field on the spacetime geometry as $\gamma$ increases.

\subsection{Varying Plasma Density}

For an inhomogeneous plasma, we use a power-law model for the plasma frequency \cite{rogers2015frequency,er2018two}:
\begin{equation}
\omega_p^2(r) = \frac{\kappa_0}{r^\alpha}, \label{eq:power_law_plasma}
\end{equation}
with $\kappa_0$ a constant and $\alpha = 1$ to focus on key traits. Substituting into Eq.~\eqref{eq:sphere_condition} gives:
\begin{equation}
\frac{2}{f(r_p)} - \frac{2 \kappa_0}{r_p \omega_0^2} - r_p \left( \frac{f'(r_p)}{f(r_p)^2} + \frac{2 \kappa_0}{r_p^3 \omega_0^2} \right) = 0. \label{eq:variable_plasma}
\end{equation}

\begin{table}[ht!]
\centering
\begin{tabular}{|c|c|c|c|}
\hline
\(\tilde{Q}\) & \(\tilde{P}\) & \(\gamma\) & \(\omega_0\) \\
\hline
0   & 0.5 & 0 & 1.0214 \\
0   & 0.5 & 1 & 1.6021 \\
0   & 1   & 0 & 0.6285 \\
0   & 1   & 1 & 0.8725 \\
0.5 & 0   & 0 & 1.0214 \\
0.5 & 0   & 1 & 1.6021 \\
0.5 & 0.5 & 0 & 0.7778 \\
0.5 & 0.5 & 1 & 1.1665 \\
0.5 & 1   & 0 & 0.5973 \\
0.5 & 1   & 1 & 0.8017 \\
1   & 0   & 0 & 0.6285 \\
1   & 0   & 1 & 0.8725 \\
1   & 0.5 & 0 & 0.5973 \\
1   & 0.5 & 1 & 0.8017 \\
1   & 1   & 0 & 0.5556 \\
1   & 1   & 1 & 0.6844 \\
\hline
\end{tabular}
\caption{Asymptotic photon energy $\omega_0$ values solving the nonlinear plasma-modified photon orbit condition at $r_p = 3$ with $M = 1$, for different electric $\tilde{Q}$, magnetic $\tilde{P}$ charges, and damping factor $\gamma$, assuming constant plasma density $\kappa_0 = 1$. Increasing $\gamma$ increases $\omega_0$, demonstrating that the nonlinearity weakens electromagnetic confinement by reducing the influence of the field on the geometry.}\label{inhomo}
\end{table}

Table~\ref{inhomo} presents the values of the asymptotic photon energy $\omega_0$ required to sustain circular orbits at $r_p = 3$ in a nonlinear EDM background with constant plasma density $\kappa_0 = 1$, under varying electric charge $\tilde{Q}$, magnetic charge $\tilde{P}$, and the damping parameter $\gamma$. These results reveal how nonlinear corrections, encoded via the exponential damping term $e^{-\gamma}$, significantly affect the electromagnetic backreaction on the photon trajectories in curved spacetime.

\section{Quantum Effects on BH Thermodynamics in EDM Fields} \label{iz9}

BHs exhibit thermodynamic properties similar to those of classical systems, governed by the principles of BH thermodynamics \cite{kubizvnak2017black,bardeen1973four,kastor2009enthalpy}. The classical Bekenstein-Hawking entropy for large BHs scales with the event horizon area:
\begin{equation}
S_0 = \pi r_h^2, \label{eq:entropy_classical}
\end{equation}
where $r_h$ is the event horizon radius. As BHs shrink via Hawking radiation, quantum corrections become significant, especially at small scales. These corrections, aligned with the holographic principle, introduce logarithmic and exponential terms to the entropy \cite{sucu2025exploring}.

To derive these, consider the microstate count $\Omega$ for a system of $N$ particles \cite{chatterjee2020exponential}:
\begin{equation}
\Omega = \frac{\left( \sum n_i \right)!}{\prod n_i!},
\end{equation}
with total particle number $N = \sum s_i n_i$ and energy $E = \sum s_i n_i \epsilon_i$, where $n_i$ is the occupancy of state $i$ with energy $\epsilon_i$. Using Stirling's approximation for large $N$, $\ln N! \approx N \ln N - N$, the most probable distribution is:
\begin{equation}
s_i = \left( \sum n_i \right) e^{-\lambda n_i},
\end{equation}
where $\lambda$ satisfies $\sum e^{-\lambda n_i} = 1$. This yields a quantum exponential correction to the entropy \cite{sucu2025exploring}:
\begin{equation}
S = S_0 + e^{-S_0}. \label{eq:entropy_corrected}
\end{equation}
When we expand the Taylor series, the modified entropy becomes:
\begin{equation}
    S\approx1+S_0^2. \label{entropy_modifiye}
\end{equation}
The corrected internal energy is:
\begin{equation}
E = \int T_H dS, \label{eq:energy}
\end{equation}
where $T_H$ is the Hawking temperature (Eq.~\eqref{hawking_modmax}). Using Eq.~\eqref{entropy_modifiye}, we get:
\begin{equation}
E \approx \frac{\pi M \,r_h^{2}}{2}-\frac{\pi \tilde{P}^{2} r_h}{{\mathrm e}^{\gamma}}-\frac{\pi \tilde{Q}^{2} r_h}{{\mathrm e}^{\gamma}}. \label{eq:energy_corrected}
\end{equation}

\begin{figure}[ht!]
    \centering
    \includegraphics[width=0.55\textwidth]{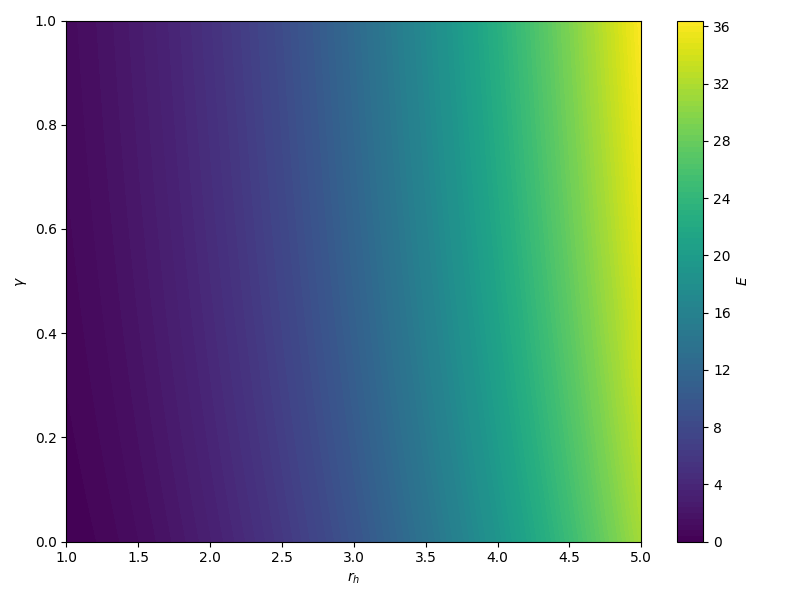}
    \caption{Density plot of the BH energy $E$ as a function of the event horizon radius $r_h$ and the ModMax parameter $\gamma$, for fixed parameters $M = 1.0$, $\tilde{Q} = 0.5$, and $\tilde{P} = 0.5$. The energy increases predominantly with $r_h$, while $\gamma$ has a more limited but monotonic effect, indicating that the BH energy is primarily sensitive to its geometric properties.}
    \label{eeee}
\end{figure}

Figure~\ref{eeee} demonstrates the relationship between the BH energy and the event horizon radius $r_h$ and the ModMax parameter $\gamma$ in the context of EDM theory. The magnitude of the energy increases significantly with $r_h$, while the change in $\gamma$ has a more limited but monotonic effect on the energy \cite{turakhonov2024weak}.

The corrected Helmholtz free energy is:
\begin{equation}
F_{EC} = -\int S_{EC} dT_H. \label{eq:helmholtz}
\end{equation}
Substituting Eq.~\eqref{entropy_modifiye} and $T_H$, we obtain:
\begin{equation}
F \approx \frac{\pi M r_h^{2}}{4}-\frac{3 \pi \tilde{P}^{2} r_h}{4 {\mathrm e}^{\gamma}}-\frac{3 \pi \tilde{Q}^{2} r_h}{4 {\mathrm e}^{\gamma}}-\frac{M}{2 \pi r_h^{2}}+\frac{\tilde{P}^{2}}{2 \pi {\mathrm e}^{\gamma} r_h^{3}}+\frac{\tilde{Q}^{2}}{2 \pi {\mathrm e}^{\gamma} r_h^{3}}.
\end{equation}

\begin{figure}[ht!]
    \centering
    \includegraphics[width=0.55\textwidth]{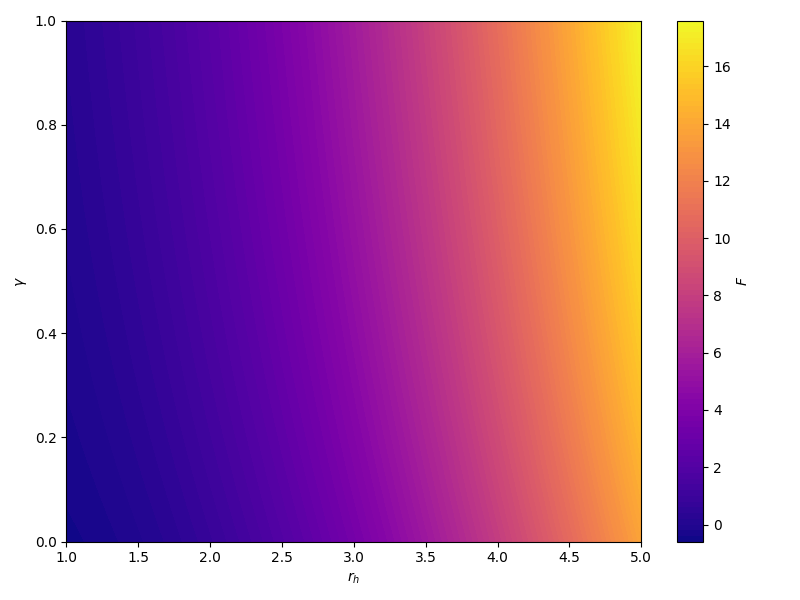}
    \caption{Density plot of the Helmholtz free energy $F$ as a function of $r_h$ and $\gamma$, for $M = 1.0$, $\tilde{Q} = 0.5$, $\tilde{P} = 0.5$. $F$ increases monotonically with both $r_h$ and $\gamma$, indicating enhanced thermodynamic energy contributions in the presence of stronger NLE effects.}
    \label{fffffff}
\end{figure}

The corrected pressure is:
\begin{equation}
P = -\frac{d F}{d V}, \label{eq:pressure}
\end{equation}
yielding:
\begin{equation}
P \approx -\frac{M}{8 r_h}-\frac{M}{4 \pi^{2} r_h^{5}}+\frac{3 \tilde{P}^{2}}{16 {\mathrm e}^{\gamma} r_h^{2}}+\frac{3 \tilde{Q}^{2}}{16 {\mathrm e}^{\gamma} r_h^{2}}+\frac{3 \tilde{P}^{2}}{8 \pi^{2} {\mathrm e}^{\gamma} r_h^{6}}+\frac{3 \tilde{Q}^{2}}{8 \pi^{2} {\mathrm e}^{\gamma} r_h^{6}}.
\end{equation}

\begin{figure}[ht!]
    \centering
    \includegraphics[width=0.55\textwidth]{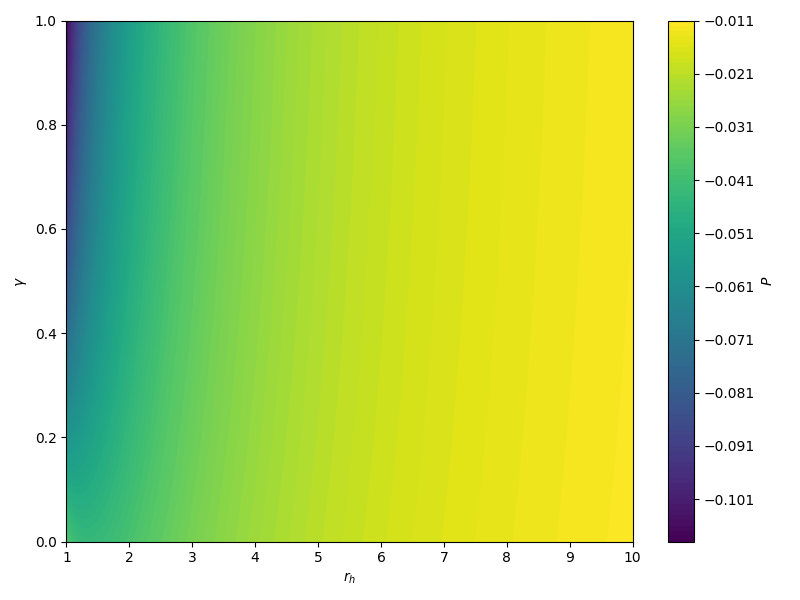}
    \caption{Density plot of the thermodynamic pressure $P$ as a function of $r_h$ and $\gamma$, for $M = 1.0$, $\tilde{Q} = 0.5$, $\tilde{P} = 0.5$. The pressure remains negative throughout the parameter space, indicating a vacuum-like equation of state. Increasing $r_h$ leads to a less negative pressure, while increasing $\gamma$ deepens the negative pressure, reflecting enhanced internal tension.}
    \label{PPPPP}
\end{figure}

The corrected heat capacity is:
\begin{equation}
C = T_H \left( \frac{\partial S}{\partial T_H} \right). \label{eq:heat_capacity}
\end{equation}
Using Eq.~\eqref{entropy_modifiye} and $T_H$, we find:
\begin{equation}
C_{EC} \approx -\frac{2 \pi^{2} r_h^{4} \left(M {\mathrm e}^{\gamma} r_h-\tilde{P}^{2}-\tilde{Q}^{2}\right)}{2 M {\mathrm e}^{\gamma} r_h-3 \tilde{P}^{2}-3 \tilde{Q}^{2}}.
\end{equation}

The denominator $2Me^{\gamma}r_h - 3(\tilde{P}^2 + \tilde{Q}^2)$ vanishes at a critical horizon radius
\begin{equation}
r_h^{*} = \frac{3(\tilde{P}^2 + \tilde{Q}^2)}{2Me^{\gamma}}\,. \label{eq:critical_rh}
\end{equation}
Since $r_h^{*}$ decreases with increasing $\gamma$, the phase transition point shifts to smaller BHs as the nonlinearity grows.

\begin{figure}[ht!]
    \centering
    \includegraphics[width=0.55\textwidth]{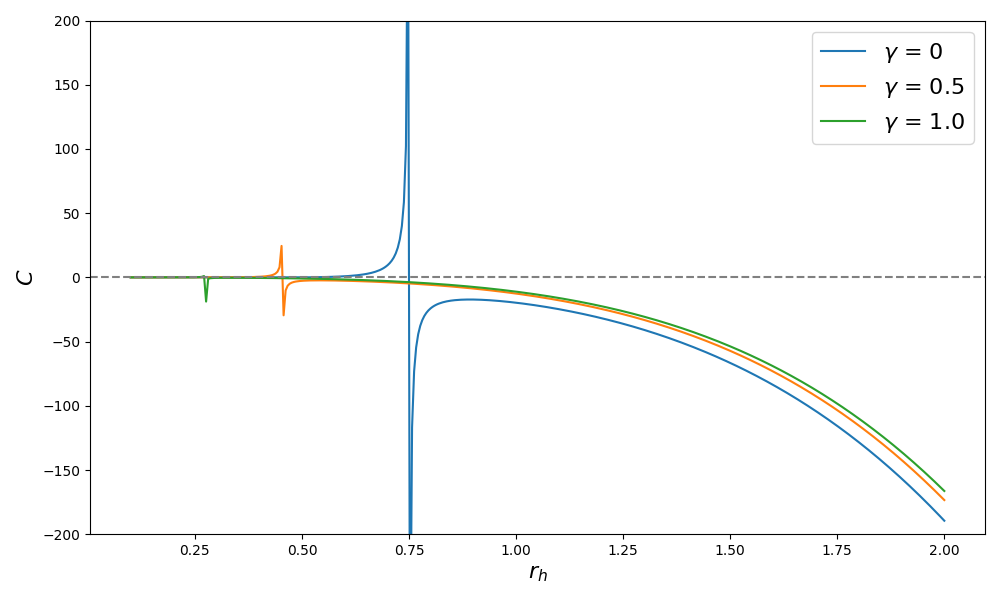}
    \caption{Heat capacity profile for EDM BH with parameters $M = 1.0$, $\tilde{Q} = 0.5$, $\tilde{P} = 0.5$, showing the temperature dependence and critical behavior near phase transitions.}
    \label{heatt}
\end{figure}

\begin{figure}[ht!]
    \centering
    \includegraphics[width=0.55\textwidth]{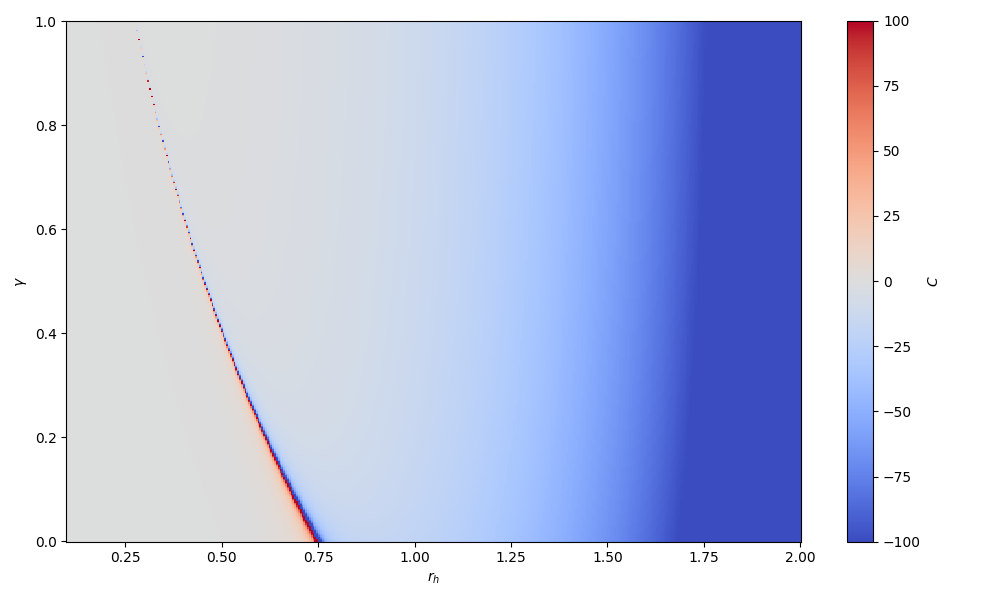}
    \caption{Heat capacity $C$ in terms of the event horizon radius $r_h$ and the modification parameter $\gamma$ by fixing the parameters $M = 1.0$, $\tilde{Q} = 0.5$, $\tilde{P} = 0.5$ within the scope of EDM theory. The color scale shows the $C$ values that determine the thermodynamic stability of the system; red regions represent positive heat capacity (stable phase), blue regions represent negative heat capacity (unstable phase). The phase transition is observed along a critical transition line given by Eq.~(\ref{eq:critical_rh}), which shifts to smaller $r_h$ values as $\gamma$ increases, revealing the regulatory effect of the EDM parameter on phase transitions.}
    \label{heatt2}
\end{figure}

Figure~\ref{heatt2} examines the heat capacity $C$, which determines the thermal stability of a charged BH defined within the scope of EDM theory, in terms of the event horizon radius $r_h$ and the modification parameter $\gamma$. The most important feature is the observation of a sharp phase transition near small $\gamma$ and a critical $r_h$ (approximately $r_h \sim 0.7$). The heat capacity suddenly changes from positive to negative along this critical line, indicating a second-order phase transition \cite{kubizvnak2012p}. This transition line shifts to smaller $r_h$ values as $\gamma$ increases, demonstrating that EDM modifications significantly affect the thermodynamic balance in regions close to the horizon.

\section{Energy Conditions and Stress-Energy Tensor Analysis} \label{iz10}
Before presenting the explicit energy condition inequalities, it is important to clarify their physical origin. At the classical level, the ModMax theory is a non-linear electrodynamics model that preserves conformal and duality symmetries and satisfies the null and weak energy conditions for physically admissible field configurations. In the present work, violations of the energy conditions appear only after the inclusion of quantum-corrected entropy contributions, indicating that these violations are not intrinsic to the classical dyonic ModMax geometry but rather reflect effective quantum modifications near the horizon. To investigate the physical viability of the spherically symmetric EDM BH spacetime under consideration, we analyze its energy-momentum tensor and the associated energy conditions \cite{dymnikova2021image,balart2014regular}. The metric function is given by:
\begin{equation}
f(r) = 1 - \frac{2M}{r} + \frac{A}{r^2}, \quad \text{with} \quad A = (\tilde{Q}^2 + \tilde{P}^2)e^{-\gamma},
\end{equation}
where $M$ denotes the BH mass, $\tilde{Q}$ and $\tilde{P}$ represent the electric and magnetic charges, respectively, and $\gamma$ is a damping parameter arising from NLE or quantum corrections.

The metric is static and spherically symmetric, allowing us to express the line element in \eqref{metric}. The corresponding Einstein tensor components for this metric are:
\begin{equation}
G^t_t = G^r_r = \frac{f'(r)}{r} + \frac{f(r) - 1}{r^2}, \quad G^\theta_\theta = G^\phi_\phi = \frac{f''(r)}{2} + \frac{f'(r)}{2r}.
\end{equation}
From Einstein's field equations $G_{\mu\nu} = 8\pi T_{\mu\nu}$, the non-zero components of the energy-momentum tensor take the form:
\begin{equation}
T^\mu_\nu = \text{diag}(-\rho,\, p_r, \,p_t, \,p_t).
\end{equation}
Computing the first and second derivatives of the metric function:
\begin{equation}
f'(r) = \frac{2M}{r^2} - \frac{2A}{r^3}, \quad f''(r) = -\frac{4M}{r^3} + \frac{6A}{r^4},
\end{equation}
we obtain explicit expressions for the energy density and pressures \cite{balart2014regular}:
\begin{equation}
\rho = -T^t_t = \frac{1}{8\pi} \left( \frac{2M}{r^3} - \frac{2A}{r^4} \right), \quad p_r = T^r_r = -\rho,
\end{equation}

\begin{equation}
p_t = T^\theta_\theta = \frac{1}{8\pi} \left( -\frac{M}{r^3} + \frac{2A}{r^4} \right).
\end{equation}
With these expressions, we proceed to examine the standard energy conditions, which serve as fundamental tests for the physical reasonableness of the spacetime \cite{visser1995lorentzian,Tangphati:2025gnq}.
\vskip 0.2 cm
\textbf{NEC:}
\begin{equation}
  \rho + p_r = 0, \quad \rho + p_t = \frac{1}{8\pi} \left( \frac{M}{r^3} \right) > 0.
\end{equation}

The second inequality is satisfied for positive mass, while the first is saturated but not strictly positive, indicating marginal compliance with this condition.
\vskip 0.2 cm
\textbf{WEC:}
\begin{equation}
  \rho \geq 0 \quad \Rightarrow \quad \frac{2M}{r^3} \geq \frac{2A}{r^4} \quad \Rightarrow \quad r \geq \frac{A}{M}.
\end{equation}

Thus, the WEC is satisfied only for radial coordinates larger than a critical radius $r_c = A/M$, and is violated near the core. This violation becomes more pronounced for larger charges or smaller $\gamma$ values, as the exponential damping factor $e^{-\gamma}$ reduces the effective charge contribution.
\vskip 0.2 cm
\textbf{SEC:}
\begin{equation}
  \rho + p_r + 2p_t = 2p_t = \frac{1}{4\pi} \left( -\frac{M}{r^3} + \frac{2A}{r^4} \right).
\end{equation}

This condition is violated for $r \lesssim 2A/M$, depending on the balance between mass and charge terms. The ModMax parameter $\gamma$ plays a crucial role here, as larger $\gamma$ values suppress the charge effects and reduce the region of SEC violation.
\vskip 0.2 cm
\textbf{DEC:}
\begin{equation}
  \rho \geq |p_r| \quad \Rightarrow \quad \rho = |p_r| \quad \text{(equality)}, \quad \rho \geq |p_t| \quad \text{depends on} \quad r.
\end{equation}

While the equality $\rho = |p_r|$ is satisfied trivially, the condition $\rho \geq |p_t|$ is generally violated at small $r$ and may hold at large $r$ \cite{visser1995lorentzian,curiel2017primer,gao2001physical}.

The analysis reveals that the ModMax parameter $\gamma$ significantly influences the energy condition violations. As $\gamma$ increases, the exponential damping factor $e^{-\gamma}$ reduces the effective electromagnetic contribution, thereby shrinking the regions where energy conditions are violated. This suggests that higher nonlinearity in the ModMax theory leads to spacetimes that are more compatible with classical energy conditions, at least in the exterior regions.

We conclude that although the energy density remains positive beyond certain critical radii, several energy conditions are violated in the near-horizon or deep interior regions of the EDM spacetime \cite{vrba2019particle}. These violations are commonly associated with exotic or quantum-corrected fields and indicate the non-classical nature of the stress-energy content supporting the BH geometry.

\section{Conclusion} \label{iz11}

In this investigation, we have explored the thermodynamic and optical properties of static, spherically symmetric EDM BHs in the presence of quantum gravity corrections and dispersive media. By extending the standard Einstein-Maxwell theory through the NLE ModMax framework, we have incorporated both electric and magnetic charges in a duality-invariant formalism, modulated by the nonlinearity parameter $\gamma$. The resulting geometry not only modifies the horizon structure but also introduces exponential damping terms that regularize the behavior of the electromagnetic field in strong-curvature regions.

Our analysis began with the fundamental EDM BH solutions, where we demonstrated how the ModMax parameter $\gamma$ controls the transition between classical Reissner-Nordstr\"{o}m-like behavior and highly nonlinear regimes. Table~\ref{tab:dyonic_modmax} catalogued the horizon structures for various parameter combinations, revealing the emergence of extremal and naked singularity configurations as functions of $\gamma$, $\tilde{Q}$, and $\tilde{P}$. The embedding diagrams in Figure~\ref{fig:dyonic_embeddings} provided intuitive visualizations of how increasing charges and nonlinearity affect the spacetime curvature.

We analyzed the Hawking radiation spectrum of EDM BHs using the Hamilton-Jacobi method and demonstrated how the temperature is affected by both dyonic charges and nonlinear corrections. Figure~\ref{hawking} illustrated the temperature profiles as functions of horizon radius for different $\gamma$ values. The incorporation of GUP revealed a natural suppression mechanism for the Hawking temperature at small scales, suggesting the possible existence of thermodynamically stable BH remnants \cite{myung2007black,bhandari2025generalized}.

On the gravitational lensing front, we employed the GBT in optical geometry to evaluate the deflection angle of light in both vacuum and plasma environments. The results demonstrate a strong dependence on the nonlinear parameter $\gamma$, as well as on plasma frequency profiles and axion-photon coupling terms in axion-plasmon scenarios. Figure~\ref{plasma1} revealed that strong lensing occurs at small impact parameters and high plasma densities \cite{bisnovatyi2010gravitational}.

The weak deflection analysis presented in Figure~\ref{lensign} demonstrated how the ModMax parameter $\gamma$ exhibits a suppressive effect on optical properties. Furthermore, the axion-plasmon environments explored through Figures~\ref{sss} and \ref{lensign_axion} showcased the combined gravitational, electromagnetic, and axionic effects on the deflection angle as the ModMax parameter $\gamma$, the plasma/reference-frequency ratio $\sigma$, the impact parameter $b$, and the axion-plasmon parameters vary \cite{pantig2022shadow,turakhonov2024weak,atamurotov2021axion}.

Our investigation of photon motion in ModMax BHs provided detailed analyses of plasma effects on photon trajectories. Tables~\ref{homojen} and~\ref{inhomo} documented how plasma frequencies and asymptotic photon energies vary with EDM parameters.

Moreover, we computed a series of quantum-corrected thermodynamic quantities: internal energy, Helmholtz free energy, pressure, and heat capacity, derived from an exponentially corrected entropy model. The density plots in Figures~\ref{eeee},~\ref{fffffff}, and~\ref{PPPPP} revealed how these thermodynamic quantities depend on both the horizon radius and the ModMax parameter. In particular, the heat capacity analysis in Figures~\ref{heatt} and~\ref{heatt2} exhibited discontinuities characteristic of second-order phase transitions, with critical points shifting as a function of $\gamma$.

The energy condition analysis demonstrated that while the exterior geometry remains observationally consistent with general relativity, the core structure involves effective fields beyond classical Einstein-Maxwell theory.

Overall, our findings illuminate the multifaceted consequences of NLE and quantum corrections on BH physics. The EDM framework, augmented by GUP and plasma effects, provides a fertile ground for modeling realistic BHs with observationally relevant signatures. Future work may include extending these results to rotating or higher-dimensional ModMax solutions, incorporating backreaction from quantum fields, or probing the observational viability of EDM lensing in strong-field astrophysical systems such as magnetars or active galactic nuclei \cite{sekhmani2025thermodynamics,okyay2022nonlinear,tang2024weak}. The present plasma lensing analysis can be naturally extended by incorporating recent developments in BH plasma optics, including non-homogeneous plasma profiles and rotating backgrounds. Such extensions may further enhance the observational relevance of ModMax BHs and will be explored in future work.

\section*{Acknowledgments}

We thank the anonymous reviewers and the handling editor for their constructive comments and suggestions that helped improve the manuscript. We would also like to thank Ercan Kilicarslan and Heeseung Zoe for useful suggestions. E.S. and \.{I}.~S. extend sincere thanks to EMU, T\"{U}B\.{I}TAK, ANKOS, and SCOAP3 for their support in facilitating networking activities under COST Actions CA22113, CA21106, CA23130, CA21136, and CA23115. During the preparation of this work, the authors used ChatGPT to refine language, improve grammar, and help with the \textit{formatting} of mathematical expressions and code \LaTeX. According\footnote{ to Elsevier's policy on the use of generative AI and AI-assisted technologies in scientific writing: \url{https://www.elsevier.com/about/policies-and-standards/generative-ai-policies-for-journals}.}  .After using this tool, the authors reviewed and edited the content as needed and take full responsibility for the content of the published article. All intellectual content, analysis, and conclusions are the authors' own.

\section*{Data Availability Statement}

This manuscript has no associated data. 

\section*{Code/Software Statement}

This manuscript has no associated code/software. [Authors' comment: Code/Software sharing not applicable as no code/software was generated during in this current study.]

\section*{Conflict of Interests}

The authors declare no conflict of interest.

\bibliographystyle{unsrt}
\bibliography{ref}

\end{document}